\documentclass[a4paper]{spie}

\usepackage{amsmath}
\usepackage{amssymb}
\usepackage{graphicx}

\usepackage{color}

\title{Electrical noise properties in aging materials}
\author{L. Buisson, M. Ciccotti, L. Bellon, S. Ciliberto
\skiplinehalf
\'Ecole Normale Sup\'erieure de Lyon, Laboratoire de Physique,\\
C.N.R.S. UMR5672, \\ 46, All\'ee d'Italie, 69364 Lyon Cedex
07, France
}

\authorinfo{correspondence to: Ludovic.Bellon@ens-lyon.fr}

\begin{document}
\maketitle

\begin{abstract}
The electric thermal noise has been measured in two aging materials, a colloidal suspension (Laponite) and a polymer (polycarbonate), presenting very slow relaxation towards equilibrium. The measurements have been performed during the transition from a fluid-like to a solid-like state for the gel and after a quench for the polymer. For both materials we have observed that the electric noise is characterized by a strong intermittency, which induces a large violation of the Fluctuation Dissipation Theorem (FDT) during the aging time, and may persist for several hours at low frequency. The statistics of these intermittent signals and their dependance on the quench speed for the polymer or on sample concentration for the gel are studied. The results are in a qualitative agreement with recent models of aging, that predict an intermittent dynamics.
\end{abstract}


\section{INTRODUCTION}

Glasses and colloids are materials that play an important role in many industrial and natural processes. One of the most puzzling properties of these materials is their very slow relaxation towards equilibrium, named aging, that presents an interesting and unusual phenomenology. More specifically when a glassy system is quenched from above to below the glass transition temperature $T_g$, any response function of the material depends on the time $t_w$ elapsed from the quench\cite{Struick}. The same phenomenon may occur in colloids during the sol-gel transition from a fluid-like to a solid-like state, that may last up to several months\cite{Kroon}. For obvious reasons related to applications, aging has been mainly characterized by the study of the slow time evolution of response functions, such as the dielectric and elastic properties of these materials. It has been observed that these systems may present very complex effects, such as memory and rejuvenation\cite{Struick,Kovacs,Jonason,bellonM}, in other words their physical properties depend on the whole thermal history of the sample.

Many models and theories have been constructed in order to explain the observed phenomenology, which is not yet completely understood. These models either predict or assume very different dynamical behaviors of the systems during aging. This dynamical behavior can be directly related to the thermal noise features of these aging systems and the study of response functions alone is unable to give definitive answers on the approaches that are the most adapted to explain the aging of a specific material. Thus it is important to associate the measure of fluctuations to that of response functions. This measurement is also related to another important aspect of aging dynamics, that is the definition of an effective temperature in these systems which are weakly, but durably, out of equilibrium. Indeed recent theories\cite{Kurchan} based on the description of spin glasses by a mean field approach proposed to extend the concept of temperature using a Fluctuation Dissipation Relation (FDR) which generalizes the Fluctuation Dissipation Theorem (FDT) for a weakly out of equilibrium system (for a review see Refs.~\citenum{Mezard,Cugliandolo,Peliti}).

For all of these reasons, in recent years, the study of the thermal noise of aging materials has received a growing interest. However, in spite of the large amount of theoretical studies there are only a few experiments dedicated to this problem\cite{Grigera,Kegel,Israeloff_Nature,Bellon,BellonD,Herisson,Weeks1,Buisson,Mazoyer,Cipelletti}. The available experimental results are in some way in contradiction and they are unable to give definitive answers. Therefore, new measurements are necessary to increase our knowledge on the thermal noise properties of these aging materials. The present work is conducted in this direction, focusing on the electrical fluctuations during the aging of two very different systems: a polymer and a colloidal gel. The interesting and common feature of the thermal noise of these two materials is the presence of a strong intermittency, that slowly disappears as function of the aging time $t_w$\cite{Buisson}. The statistical features of this intermittency have a qualitative agreement with those of recent theoretical models which predict an intermittent dynamics for the local and global variables of an aging system. These experimental results and a comparison with the theoretical models are reported in the following sections of this communication. In section 2, the electrical measurements performed on the polymer and the statistical features of the thermal noise signal are described. In section 3 the properties of electric fluctuations in a colloidal suspension of Laponite are analyzed. In section 4 we compare the experimental results with the theoretical ones before concluding.

\section{POLYCARBONATE DIELECTRIC PROPERTIES AND THERMAL NOISE}

We present in this section measurements of the dielectric susceptibility and of the polarization noise, in the range $20\,mHz\,-\,100\,Hz$, of a polymer glass: polycarbonate. These results demonstrate the appearance of a strong intermittency of the noise when this material is quickly quenched from the molten state to below its glass-transition temperature. This intermittency produces a strong violation of the FDT at very low frequency. The violation is a decreasing function of time and frequency and it is still observed for $\omega t_w \gg 1$: it may last for more than $3h$ for $f>1\,Hz$. We have also observed that the intermittency is a function of the cooling rate of the sample and almost disappears after a slow quench. In this case the violation of FDT remains, but it is very small.

\subsection{Experimental setup}

The polymer used in this investigation is Makrofol DE 1-1 C, a bisphenol A polycarbonate, with $T_g \simeq 419K$, produced by Bayer in form of foils. We have chosen this material because it has a wide temperature range of strong aging\cite{Struick}. This polymer is totally amorphous: there is no evidence of crystallinity\cite{Wilkes1}. Nevertheless, the internal structure of polycarbonate changes and relaxes as a result of a change in the chain conformation by molecular motions\cite{Struick,Duval,Quinson}. Many studies of the dielectric susceptibility of this material exist, but none had an interest on the problem of noise measurements.

\begin{figure}[tb]
\begin{center}
\includegraphics[width=9.5cm]{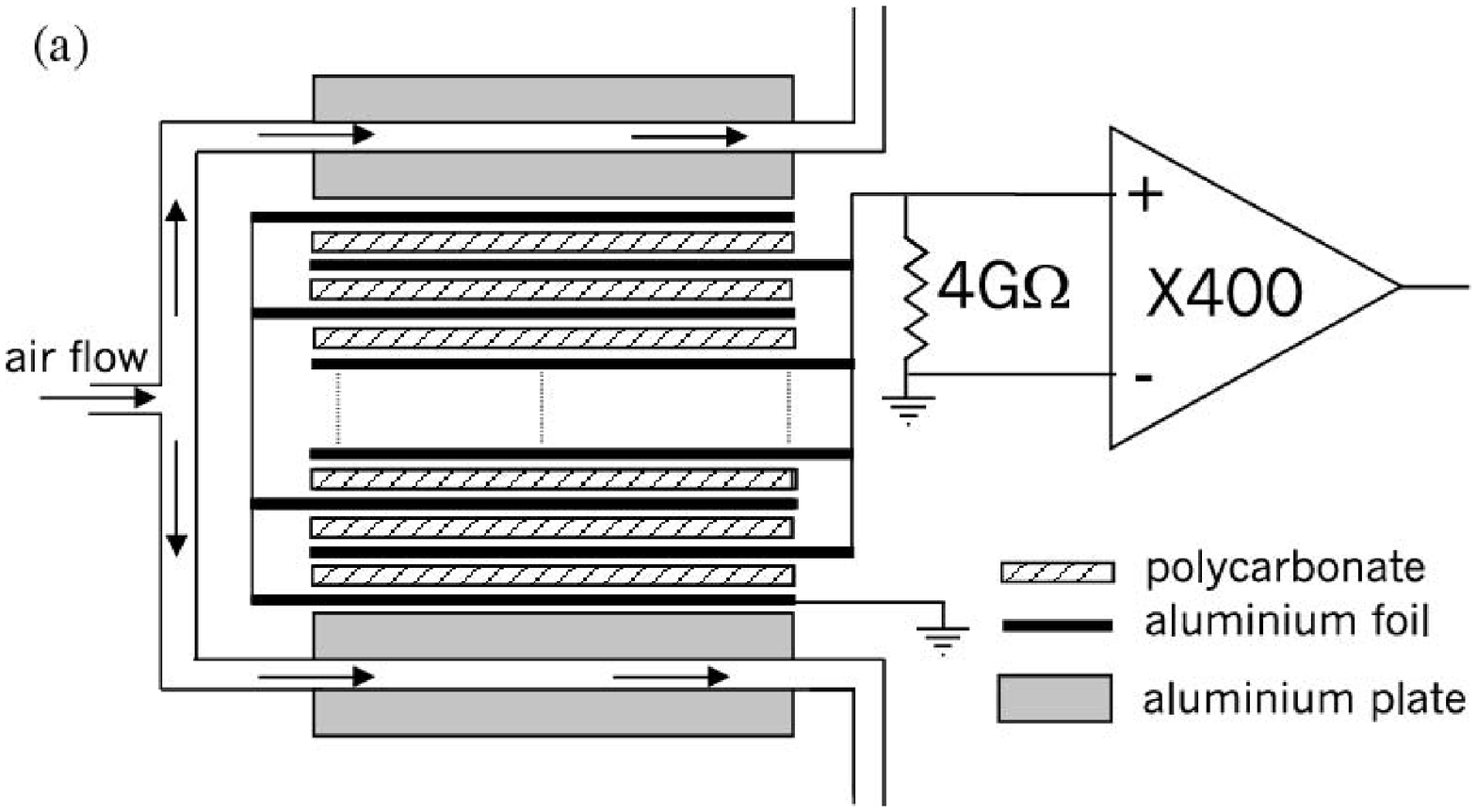}
\hspace{10mm}
\includegraphics[width=5.5cm]{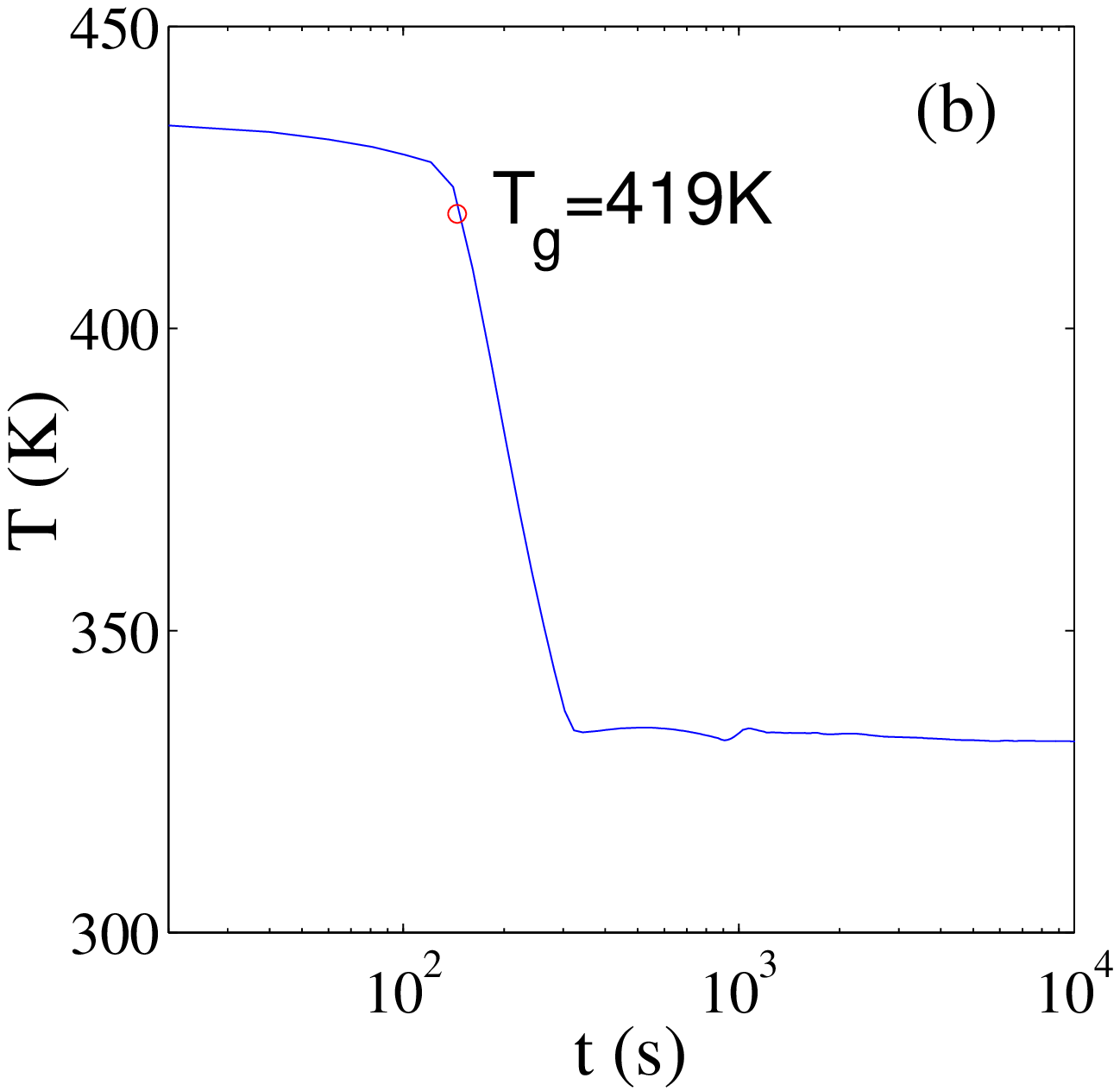}
\end{center}
\caption{{\bf Polycarbonate experimental set-up. }(a) Design of polycarbonate capacitance cell. (b) Typical temperature quench: from $T_i=453\,K$ to $T_f=333\,K$, the origin of $t_w$ is set at $T=T_g$. }
\label{Experimental set-up}
\end{figure}

In our experiment polycarbonate is used as the dielectric of a capacitor. The impedance is composed by $14$ cylindrical capacitors in parallel in order to reduce the resistance of the sample and to increase its capacity. Each capacitor is made of two aluminum electrodes, $12\,\mu m$ thick, and by a disk of polycarbonate of diameter $12\,cm$ and thickness $125\,\mu m$. The experimental set-up is shown in Fig.~\ref{Experimental set-up}(a). The $14$ capacitors are sandwiched together and hold between two thick aluminum plates which contain an air circulation used to regulate the sample temperature. This mechanical design of the capacitor is very stable and gives very reproducible results even after many temperature quenches. The capacitor is inside 4 Faraday screens to insulate it from external noise. The temperature of the sample is controlled within a few percent. Fast quenches of about $50\,K/min$ are obtained by injecting Nitrogen vapor in the air circulation of the aluminum plates. The electrical impedance of the capacitor is $Z(\omega,t_w) = R / (1+i \omega \ R \ C)$, where $C$ is the capacitance and $R$ is a parallel resistance which accounts for the complex dielectric susceptibility. This is measured by a circuit calibrated using a Novocontrol Impedance Analyzer. The noise spectrum of this impedance $S_Z(\omega,t_w)$ is\cite{Nyquist,book}:
\begin{equation}
S_Z(f,t_w)= 4 \ k_B \ T_{ef\!f}(f, t_w) \ Re [Z(\omega,t_w)]= \frac{4 \
k_B \ T_{ef\!f}(f, t_w) \ R}{1+ (\omega \ R \ C)^2 } \label{SZ}
\end{equation}
where $k_B$ is the Boltzmann constant and $T_{ef\!f}$ is the effective temperature of the sample. This effective temperature takes into account the fact that FDT (Nyquist relation for electric noise) can be violated because the polymer is out of equilibrium during aging, and in general $T_{ef\!f}>T$, with $T$ the temperature of the thermal bath\cite{Mezard}. Of course when FDT is satisfied then $T_{ef\!f}=T$. In order to measure $S_Z(f,t_w)$, we made a differential amplifier based on selected low noise JFET (2N6453 InterFET Corporation), the input of which is polarized by an input resistance $R_i= 4\,G\Omega$. Above $2\,Hz$, the input voltage noise of this amplifier is $5\,nV/\sqrt{Hz}$ and the input current noise is about $1\,fA/\sqrt{Hz}$. The output signal of the amplifier is analyzed either by an HP3562A dynamic signal analyzer or directly acquired by a NI4462 card. It is easy to show that the measured spectrum at the amplifier input is:
\begin{equation}
S_V(f,t_w) = \frac{4 \ k_B \ R \ R_i \ \ (\ T_{ef\!f}(f, t_w) \ R_i +
\ T_R \ R + S_\xi(f) \ R \ R_i )}{ (R+R_i)^2+(\omega \ R \
R_i \ C)^2} + S_{\eta}(f)
\label{Vnoise}
\end{equation}
where $T_R$ is the temperature of $R_i$ and $S_\eta$ and $S_\xi$ are respectively the voltage and the current noise spectrum of the amplifier. In order to reach the desired statistical accuracy of $S_V(f,t_w)$, we averaged the results of many experiments. In each of these experiments the sample is first heated to $T_i=1.08\,T_g$. It is maintained at this temperature for several hours in order to reinitialize its thermal history. Then it is quenched from $T_i$ to the working final temperature $T_f$ where the aging properties are studied. The minimum quenching time from $T_i$ to $T_f$ is $120\,s$. A typical thermal history of the quench is shown in Fig.~\ref{Experimental set-up}(b). The reproducibility in the measurement of the capacitor impedance, during this thermal cycle, is always better than $1\%$. The origin of aging time $t_w$ is the instant when the capacitor temperature is $T=T_g \simeq 419\,K$, which of course may depend on the cooling rate. However adjustment of $T_g$ of a few degrees will shift the time axis by at most $30\,s$, without affecting our results.

\begin{figure}[tb]
\begin{center}
\null
\hfill
\includegraphics[height=56mm]{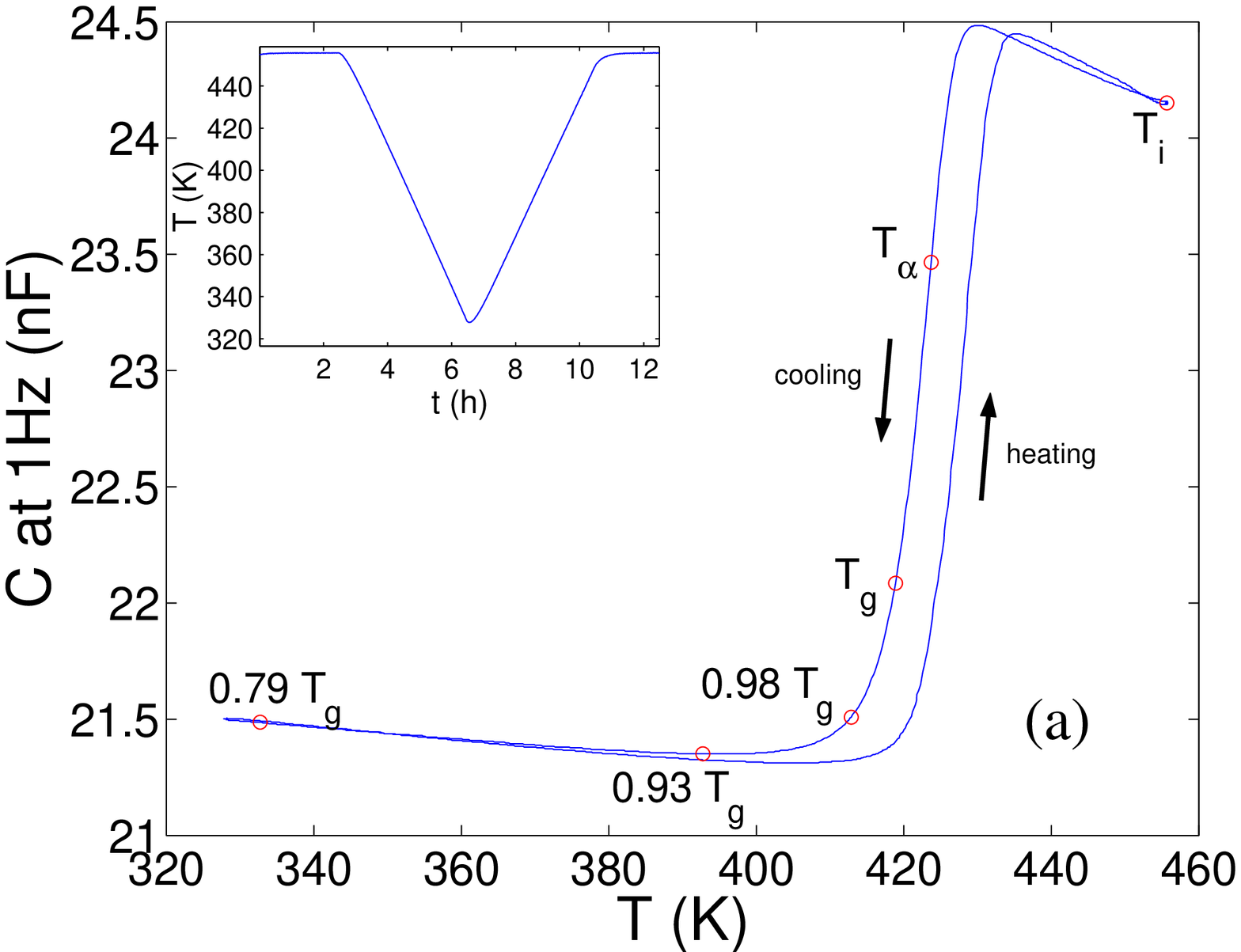}
\hfill
\includegraphics[height=56mm]{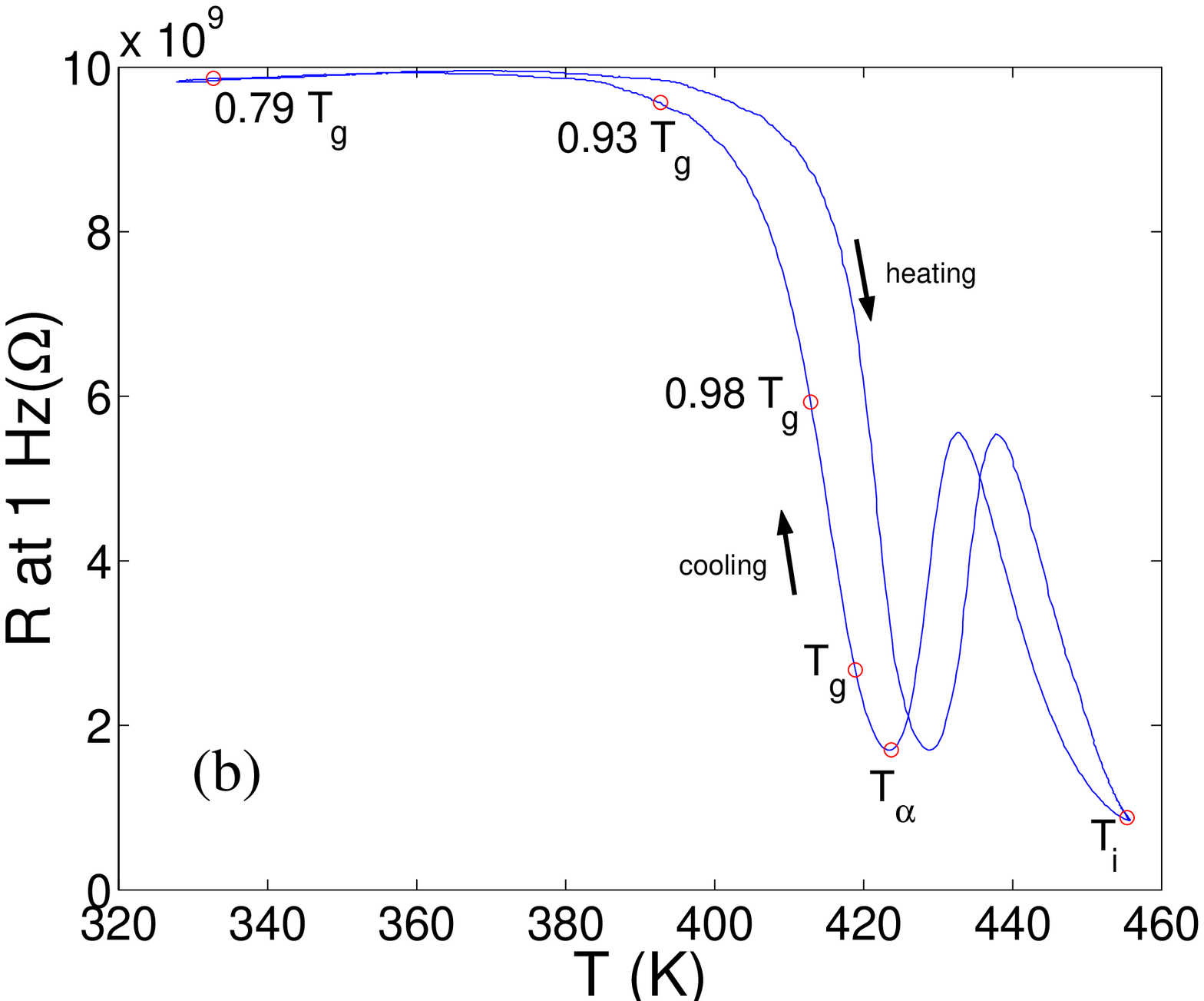}
\hfill
\null
\end{center}
 \caption{{\bf Polycarbonate response function at 1Hz during a temperature cycle. }(a) Dependence of $C$ on temperature, when $T$ is changed as function of time as indicated in the inset. (b) Dependence of $R$ on $T$. $T_\alpha$ is the temperature of the $\alpha$ relaxation at $1\,Hz$, $T_g$ is the glass transition temperature. The other circles on the curve indicate the $T_f$ where aging has been studied.}
 \label{hist}
 \end{figure}

\begin{figure}[!ht]
\begin{center}
\includegraphics[height=50mm]{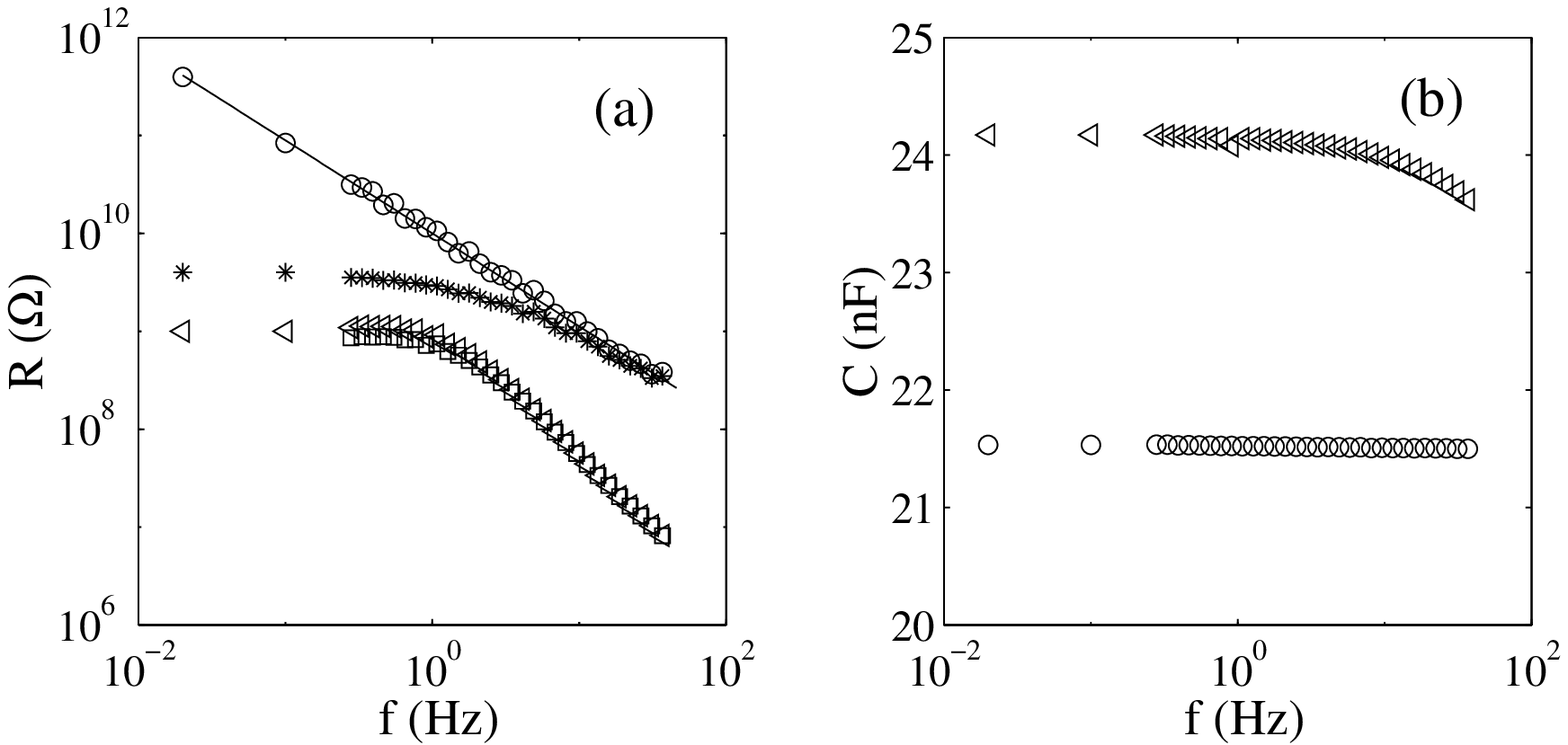}
\hspace{7mm}
\includegraphics[height=50mm]{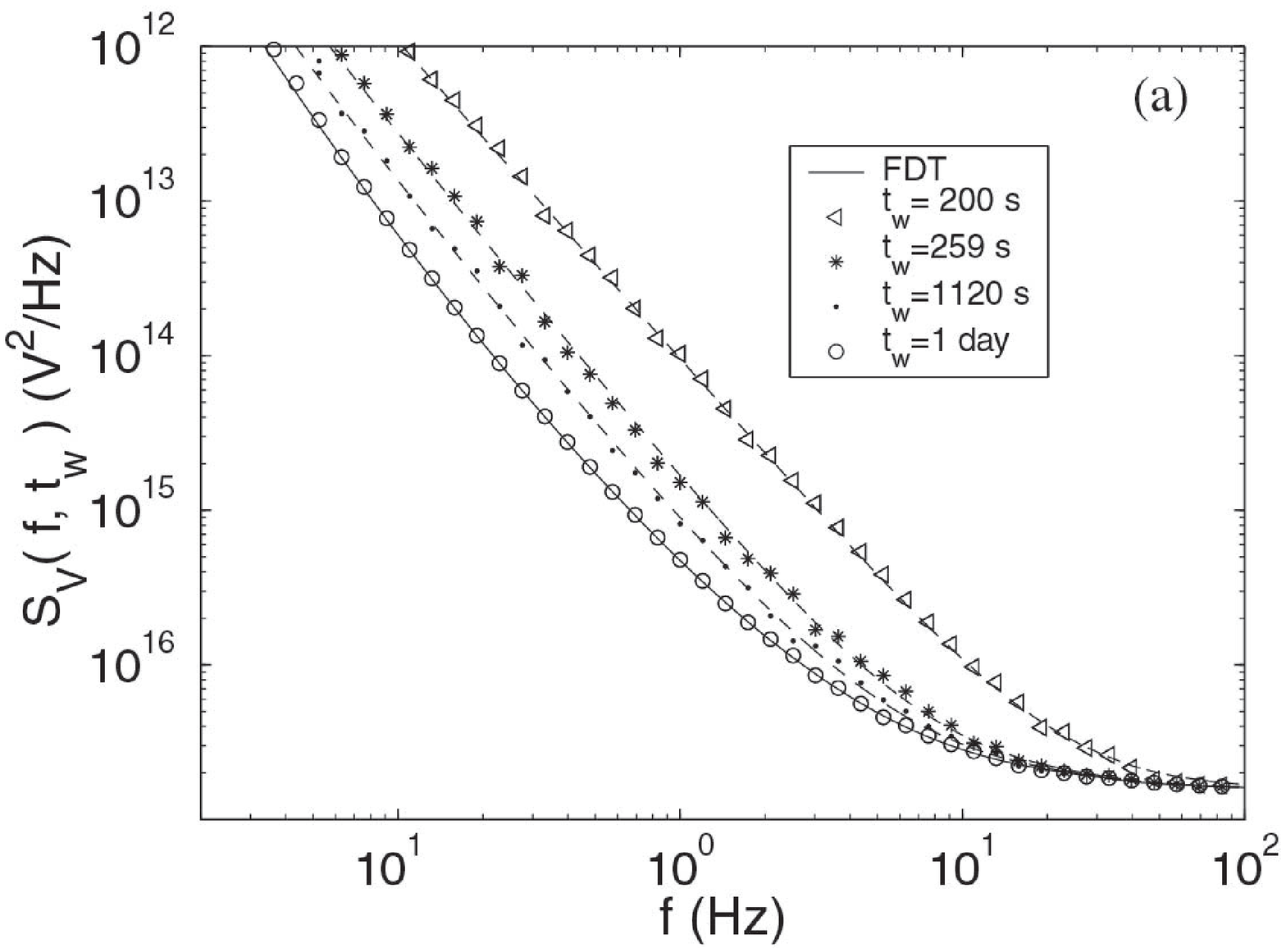}
\end{center}
\caption{{\bf Polycarbonate response function after a fast quench. }(a) Polycarbonate resistance $R$
 as a function of frequency measured at $T_i=1.08\,T_g$ ($\vartriangleleft$) and
 at $T_f=0.79\,T_g$ ($\circ$). The effective resistance at the input of the amplifier (parallel of the $4\,G\Omega$ input resistance and of the polycarbonate impedance) is also plotted at $T_i$ ($\square$) and at $T_f$ ($\ast$).
 (b) Polycarbonate capacitance versus frequency measured at $T_i $ ($\vartriangleleft$) and
 at $T_f$ ($\circ$).
(c) Typical aging of $R$ measured at $1\,Hz$ as a function of $t_w$ after a quench at $T_f=0.79\,T_g$.}
\label{reponse}
 \end{figure}


\subsection{Response and noise measurements}

Before discussing the time evolution of the dielectric properties and of the thermal noise at $T_f$ we show in Fig.~\ref{hist} the dependence of $C$ and $R$ measured at $1\,Hz$ as a function of temperature, which is ramped as a function of time as indicated in the inset of Fig.~\ref{hist}(a). We notice a strong hysteresis between cooling and heating. In the figure $T_\alpha$ is the temperature of the $\alpha$ relaxation at $1\,Hz$. The other circles on the curve indicate the $T_f$ where the aging has been studied. We performed measurements at $T_f=0.79\,T_g,0.93\,T_g$ and $0.98\,T_g$. We first describe the results at the smallest temperature, that is $T_f=0.79\,T_g$. In Figs.~\ref{reponse}(a) and (b), we plot the measured values of $R$ and $C$ as a function of frequency $f$ at $T_i=1.08\,T_g$ and at $T_f$ for $t_w \geqslant 200\,s$. The dependence of $R$, at $1\,Hz$, as a function of time is shown in Fig.~\ref{reponse}(c). We see that the time evolution of $R$ is logarithmic in time for $t>300\,s$ and that the aging is not very large at $T_f=0.79\,T_g$, only $10\%$ in the first $3\,h$. At higher temperature, close to $T_g$, aging is much larger.

Looking at Fig.~\ref{reponse}(a) and (b), we see that when lowering temperature $R$ increases and $C$ decreases. As at $0.79\,T_g$ aging is small and extremely slow for $t_w>200\,s$ the impedance can be considered constant without affecting our results. From the data plotted in Fig.~\ref{reponse} (a) and (b) one finds that $R=10^{10}(1 \pm 0.05) \ f^{-1.05\pm 0.01} \,\Omega$ and $C=(21.5 \pm 0.05)\,nF$. In Fig.~\ref{reponse}(a) we also plot the total resistance at the amplifier input which is the parallel of the capacitor impedance with $R_i$. We see that at $T_f$ the input impedance of the amplifier is negligible for $f>10\,Hz$, whereas it has to be taken into account at slower frequencies.

\begin{figure}
\begin{center}
\null
\hfill
\includegraphics[height=60mm]{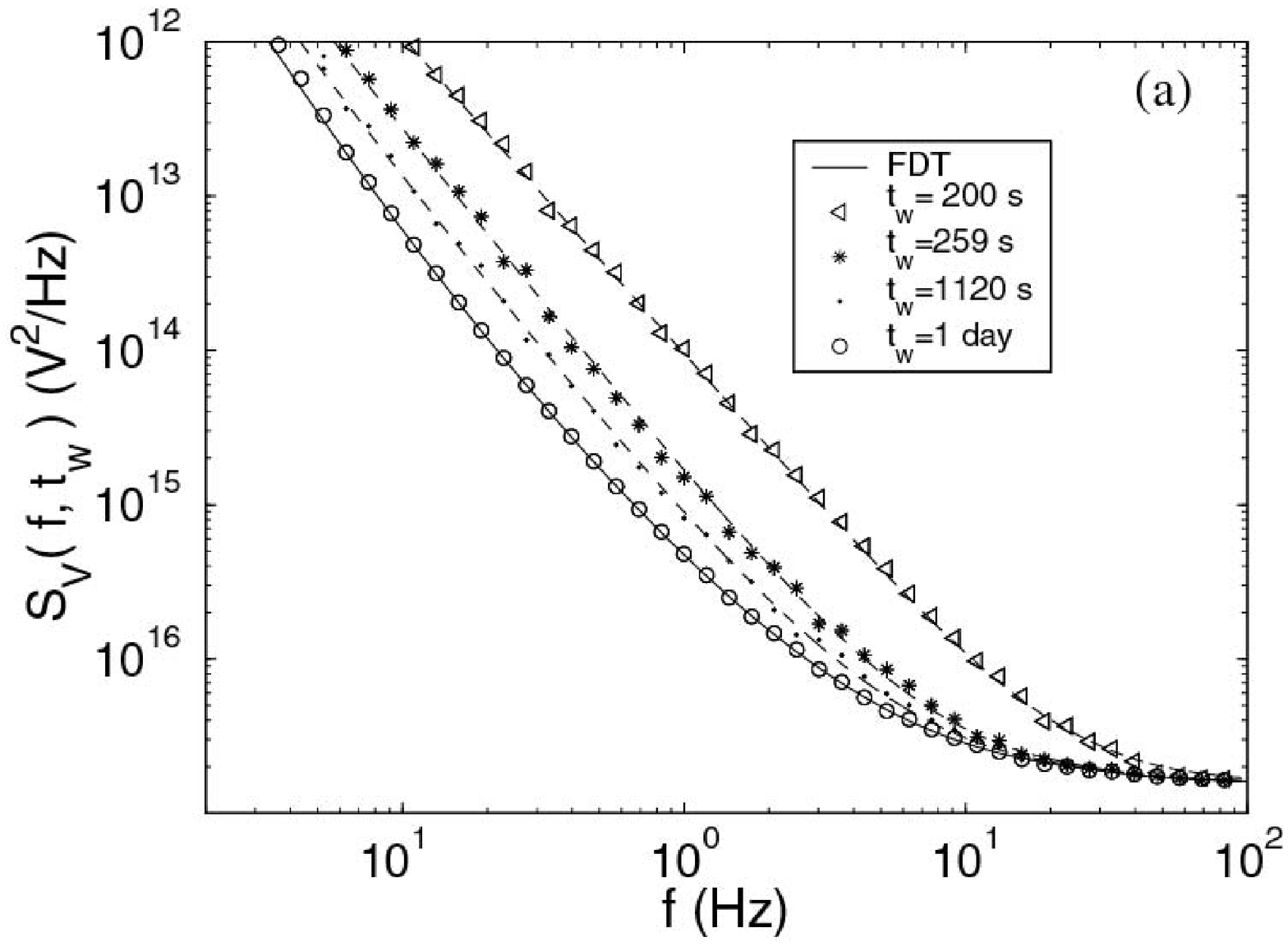}
\hfill
\includegraphics[height=60mm]{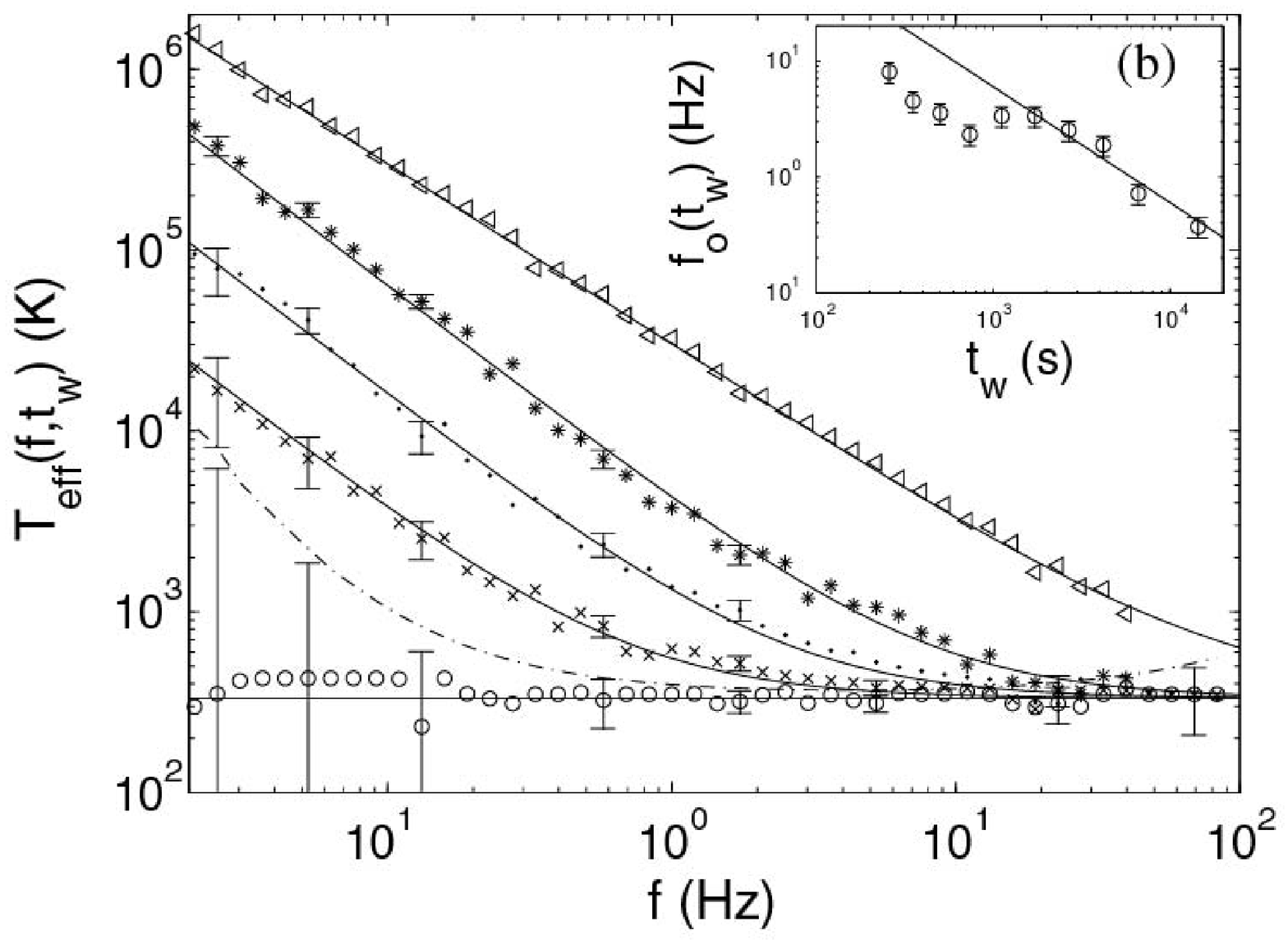}
\hfill
\null
\end{center}

\caption{{\bf Voltage noise and effective temperature in polycarbonate after a fast quench at $T_f=0.79\,T_g$. }(a) Noise power spectral density $S_V(f,t_w)$ measured at different $t_w$. The spectra are averaged over seven quenches. The continuous line is the FDT prediction. Dashed lines are the fit obtained using Eqs.~(\ref{Vnoise}) and (\ref{fitTeff}) (see text for details). (b) Effective temperature vs. for different aging times: $(\vartriangleleft)\ t_w= 200 \, s$, $ (\ast)\ t_w= 260\,s$, $ (\bullet) \ t_w= 2580\,s$, $ (\times) t_w=6542\,s$, $(\circ)\ t_w= 24\,h$. The continuous lines are the fits obtained using Eq.~(\ref{fitTeff}). The horizontal straight line is the FDT prediction. The dot dashed line corresponds to the limit where the FDT violation can be detected. In the inset the frequency $f_o(t_w)$, defined in Eq.~(\ref{fitTeff}), is plotted as a function of $t_w$. The continuous line is not a fit, but corresponds to $f_o(t_w) \propto 1/t_w$. } \label{noise}
\end{figure}

Fig.~\ref{noise}(a) represents the evolution of $S_V(f,t_w)$ after a quench. Each spectrum is obtained as an average in a time window starting at $t_w$. The time window increases with $t_w$ in order to reduce errors for large $t_w$. Then the results of 7 quenches have been averaged. For the longest time ($t_w=24\,h$), the equilibrium FDT prediction (continuous line) is quite well satisfied. However, we clearly see that FDT is strongly violated for all frequencies at short times. Then high frequencies relax on the FDT, but there is a persistence of the violation for lower frequencies. The amount of the violation can be estimated by the best fit of $T_{ef\!f}(f,t_w)$ in Eq.~(\ref{Vnoise}) where all other parameters are known. We started at very large $t_w$ when the system has relaxed and $T_{ef\!f}=T_f$ for all frequencies. Inserting the values in Eq.~(\ref{Vnoise}) and using the $S_V$ measured at $t_w=24\,h$ we find $T_{ef\!f}\simeq T_f=333\,K$, within error bars for all frequencies --- see Fig.~\ref{noise}(b). At short $t_w$, data show that $T_{ef\!f}(f,t_w)\simeq T_f$ for $f$ larger than a cutoff frequency $f_o(t_w)$ which is a function of $t_w$. In contrast, for $f<f_o(t_w)$, we find that $T_{ef\!f}(f,t_w)\propto f^{-A(t_w)}$, with $A(t_w)\simeq 1$. This frequency dependence of $T_{ef\!f}(f,t_w)$ is quite well approximated by
\begin{equation}
T_{ef\!f}(f,t_w)= T_f \ \left[ \ 1 \ + \ \left( \frac{f}{f_o(t_w)}\right)^{-A(t_w)} \ \right]
 \label{fitTeff}
\end{equation}
where $A(t_w)$ and $f_o(t_w)$ are the fitting parameters. We find that $1<A(t_w)<1.2$ for all the data set. Furthermore, for $t_w \geq 250\,s$, it is enough to set $A(t_w)=1.2$ to fit the data within error bars. For $t_w <250\,s$ we fixed $A(t_w)=1$. Thus the only free parameter in Eq.~(\ref{fitTeff}) is $f_o(t_w)$. The continuous lines in Fig.~\ref{noise}(a) are the best fits of $S_V$ found inserting Eq.~(\ref{fitTeff}) in Eq.~(\ref{Vnoise}).

In Fig.~\ref{noise}(b) we plot the estimated $T_{ef\!f}(f,t_w)$ as a function of frequency at different $t_w$. We see that just after the quench $T_{ef\!f}(f,t_w)$ is much larger than $T_f$ in all the frequency interval. High frequencies rapidly decay towards the FDT prediction whereas at the smallest frequencies $T_{ef\!f}\simeq 10^5K$. Moreover, we notice that low frequencies decay more slowly than high frequencies and that the evolution of $T_{ef\!f}(f,t_w)$ towards the equilibrium value is very slow. From the data of Fig.~\ref{noise}(b) and Eq.~(\ref{fitTeff}), it is easy to see that the $T_{ef\!f}(f,t_w)$ curves can be superposed onto a master curve by plotting them as a function of $f/f_o(t_w)$. The function $f_o(t_w)$ is a decreasing function of $t_w$, but the dependence is not a simple one, as it can be seen in the inset of Fig.~\ref{noise}(b). The continuous straight line is not fit, it represents $f_o(t_w)\propto 1/t_w$ which seems a reasonable approximation for these data. For $t_w > 10^4\,s$ we find the $f_o<1\,Hz$. Thus we cannot follow the evolution of $T_{ef\!f}$ anymore because the contribution of the experimental noise on $S_V$ is too important, as it is shown in Fig.~\ref{noise}(b) by the growth of the error bars for $t_w=24\,h$ and $f<0.1\,Hz$.

For other working temperatures in the range $0.79\,T_g<T_f<0.93\,T_g$, where the low frequency dielectric properties are almost temperature independent (see Fig.~\ref{hist}(b)), the same scenario appears. The only important difference to mention here is that aging becomes faster and more pronounced as $T_f$ increases. At $T_f=0.93\,T_g$ the
losses of the capacitor change of about $50\%$ in about $3h$, but all the spectral analyses performed after a fast quench give the same evolution. For $T_f>0.93\,T_g$ fast quenches cannot be done for technical reasons and the results are indeed quite different. Thus
we will not consider, for the moment, the measurement at $T_f=0.98\,T_g$ and we will mainly focus on the experiments performed in the range $0.79\,T_g<T_f<0.93\,T_g$ with fast quenches. For these measurements the spectral analysis on the noise signal, described in this section, indicates that Nyquist relation (FDT) is strongly violated for a long time after the quench. The question is now to understand the reasons of this violation.

\begin{figure}[tb]
\begin{center}
\null
\hfill
\includegraphics[width=7cm]{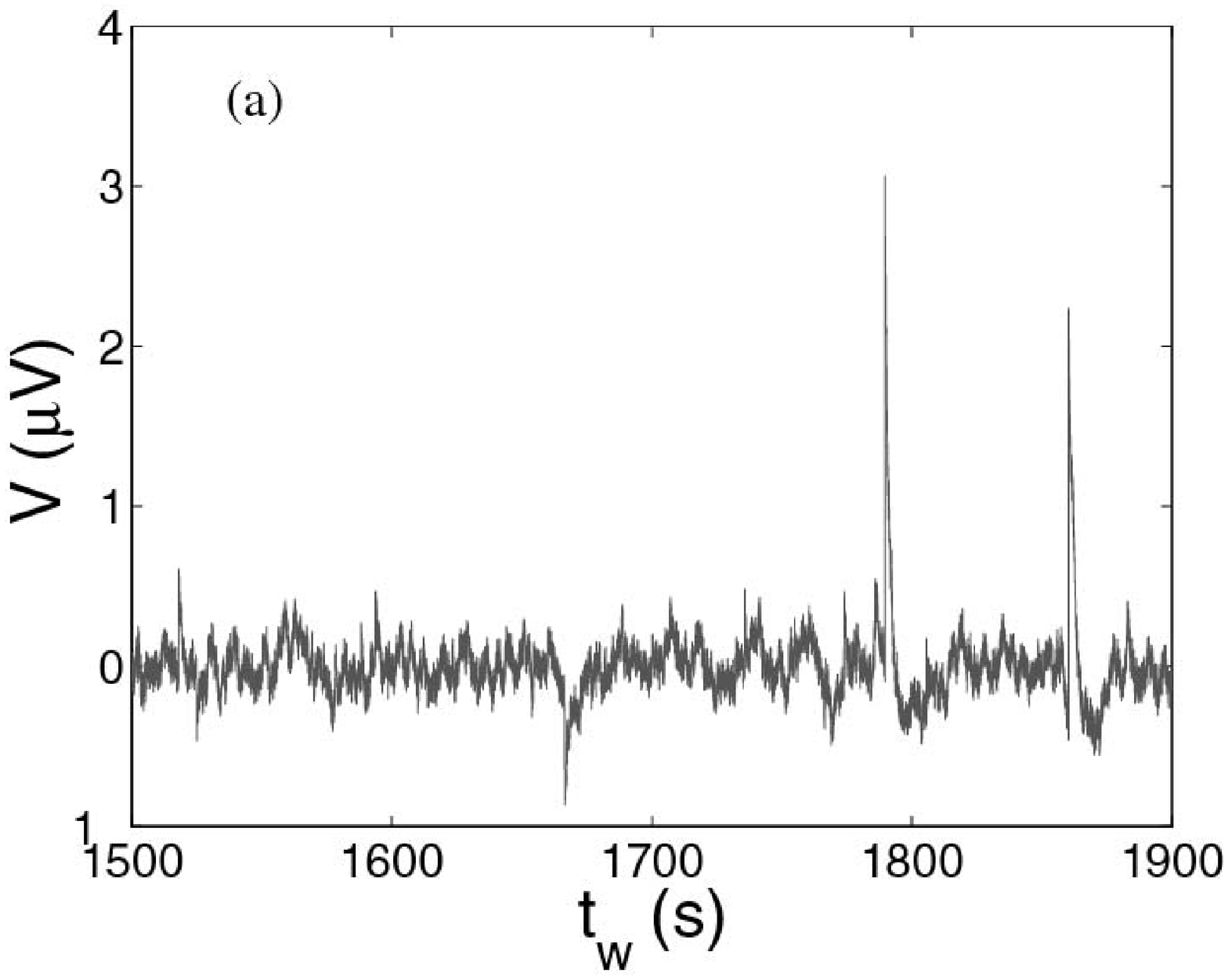}
\hfill
\includegraphics[width=7cm]{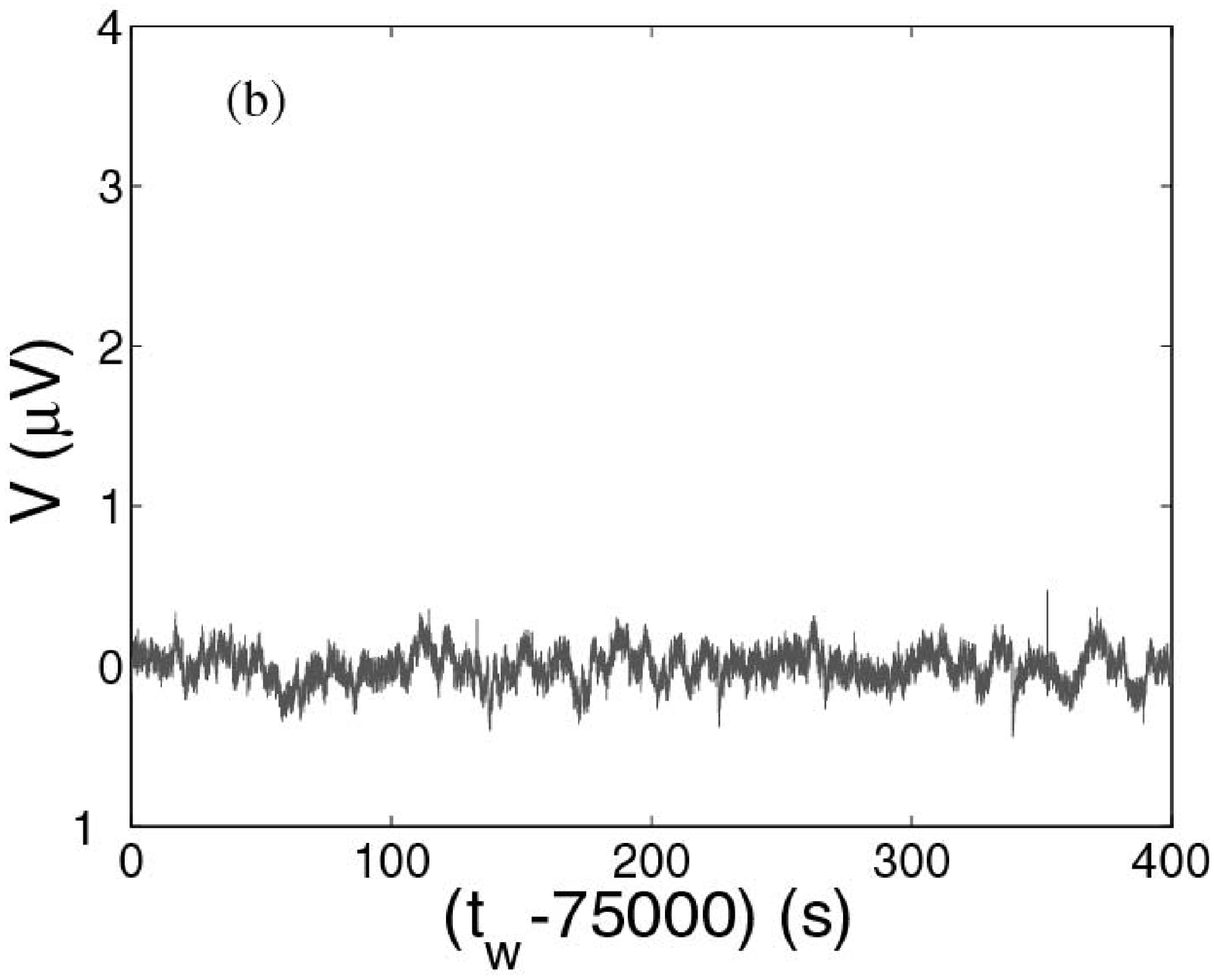}
\hfill
\null
 \end{center}
\caption{{\bf Voltage noise signal in polycarbonate after a fast quench at $T_f=0.79\,T_g$. }Typical
noise signal measured when (a) $1500\,s<t_w<1900\,s$ and (b) $t_w>75000\,s$}
 \label{signalpolyca}
\end{figure}

\begin{figure}[tb]
\centerline{ \includegraphics[width=8cm]{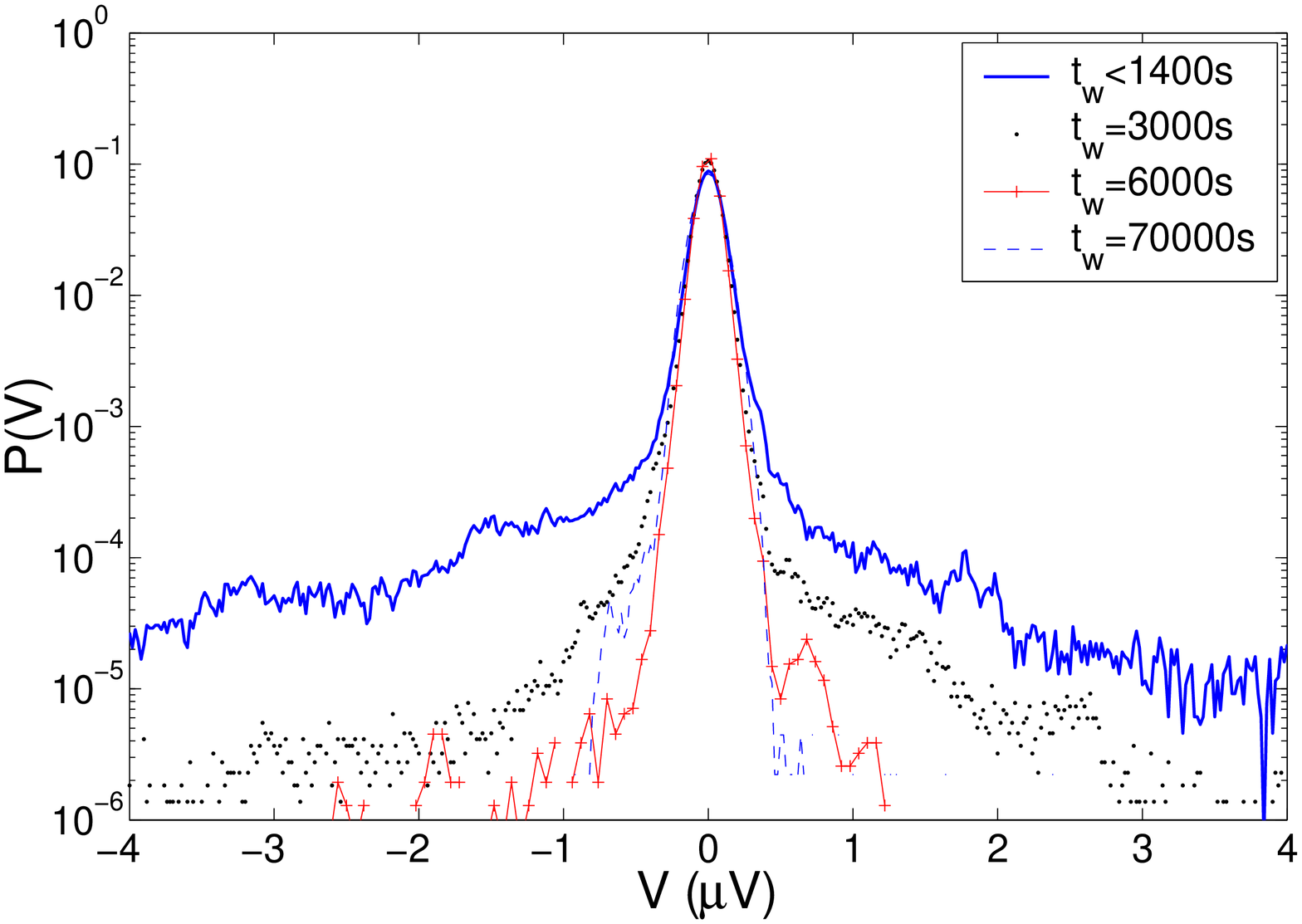}}
\caption{{\bf PDF of voltage noise in polycarbonate after a fast quench at $T_f=0.79\,T_g$. }The heavy tails of the PDF at early $t_w$ are a signature of strong intermittency. The distribution eventually relaxes on the gaussian shape of classic thermal noise.}
\label{PDFpolyca}
\end{figure}

\begin{figure}[tb]
\begin{center}
\null
\hfill
\includegraphics[width=7cm]{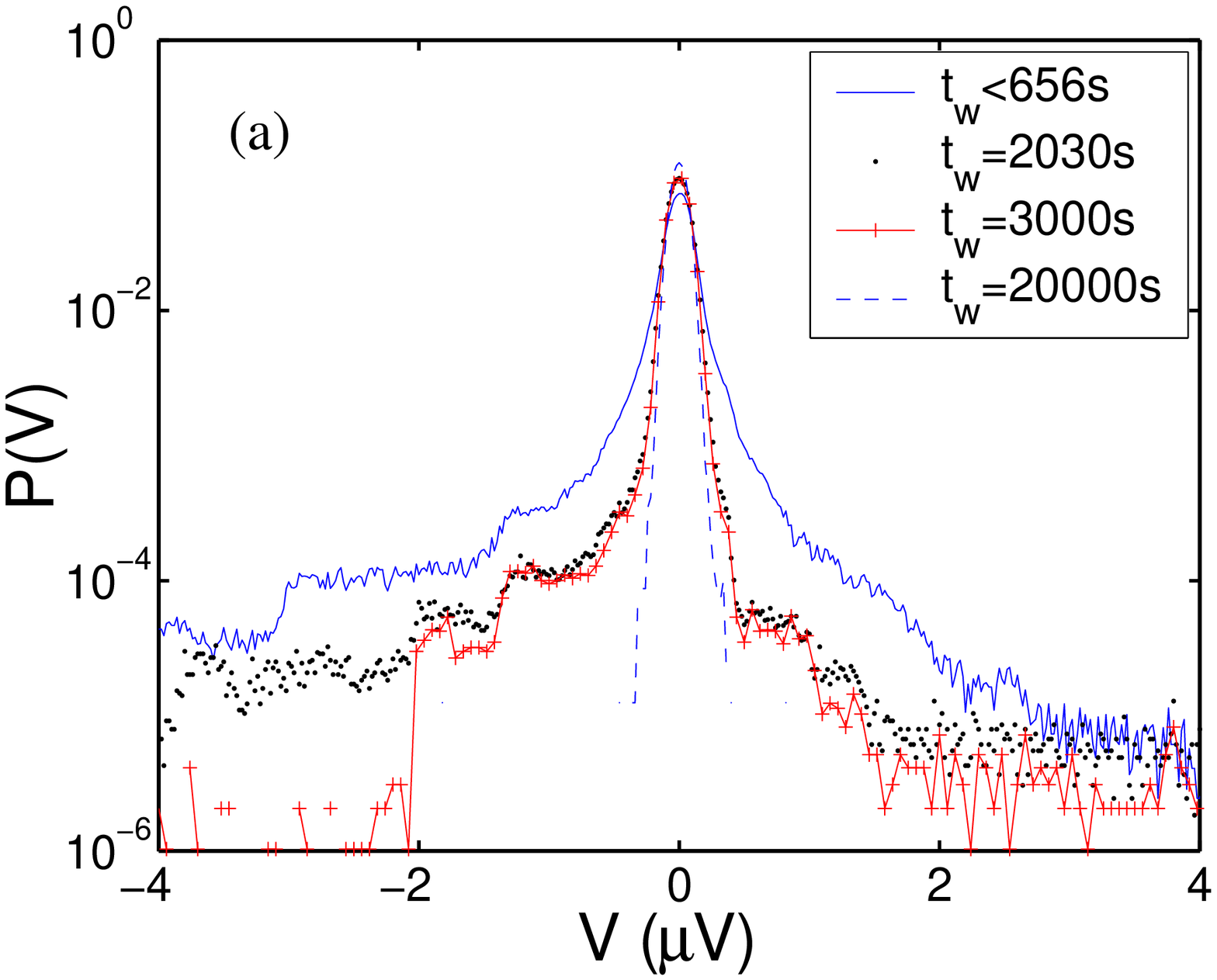}
\hfill
\includegraphics[width=7cm]{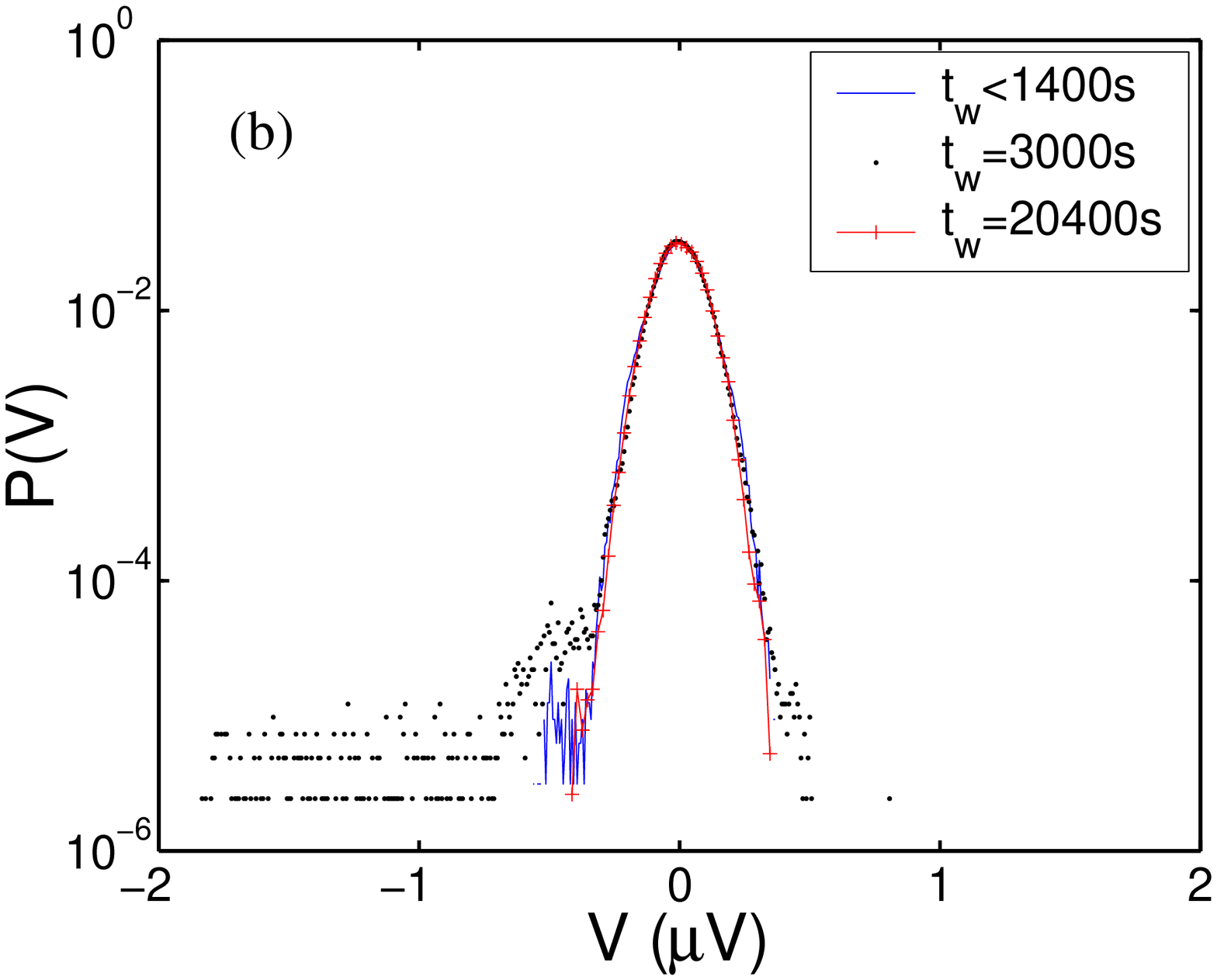}
\hfill
\null
\end{center}
\caption{{\bf PDF of voltage noise in polycarbonate after a quench at $T_f=0.93\,T_g$. } (a) Strong intermittency follows a fast quench at $50\,K/min$. (b) Almost no intermittency is visible after a slow quench at $3.6\,K/min$.}
\label{PDF120}
\end{figure}

\subsection{ Statistical analysis of the noise}

In order to understand the origin of such large deviations in our experiment we directly analyzed the noise signal. We found out that the signal is characterized by large intermittent events which produce low frequency spectra proportional to $f^{-\alpha}$ with $\alpha \simeq 2$. Two typical signals recorded at $T_f=0.79\,T_g$ for $1500\,s<t_w<1900\,s$ and $t_w>75000\,s$ are plotted in Fig.~\ref{signalpolyca}. We clearly see that in the early signal there are very large bursts which are at the origin of the frequency spectra discussed in the previous section. In contrast in the later signal of Fig.~\ref{signalpolyca}(b), which was recorded when FDT is not violated, the bursts totally disappear.

The probability density function (PDF) of these signals is shown in Fig.~\ref{PDFpolyca}. We clearly see that the PDF, measured at small $t_w$, presents very high tails which become smaller and smaller at large $t_w$. Finally the Gaussian profiles is recovered after $24h$. The same behavior is observed at $T_f=0.93\,T_g$ after a fast quench, as shown in Fig.~\ref{PDF120}(a). The only difference is the relaxation rate towards the Gaussian distribution, which is faster at higher temperature, where aging effects are larger. From these measurements one concludes that after a fast quench the electrical thermal noise is strongly intermittent, which is the origin of the large violation of the FDT.

However this behavior depends on the quench speed. In Fig.~\ref{PDF120}(b) we plot the PDF of the signals measured after a slow quench at $3.6\,K/min$. We clearly see that the PDF are very different. Intermittency has almost disappeared and the violation of FDT is now very small, about $15\%$. The comparison between the fast quench and the slow quench merits a special comment. During the fast quench $T_f=0.93\,T_g$ is reached in about $100\,s$ after the passage of $T$ at $T_g$. For the slow quench this time is about $1000\,s$. Therefore one may wonder whether after the first $1000\,s$ of the fast quenched sample, one recovers the same dynamics of the slow quenched one. By comparing the PDF of Fig.~\ref{PDF120}(a) with those of Fig.~\ref{PDF120}(b) we clearly see that this is not the case. Indeed for $t_w \simeq 1000\,s$ after the fast quench the tails of the PDF are still much larger than those of the slow quench. Therefore one deduces that the polymer is actually following a completely different dynamics according to the cooling rate. This is a very important observation that can be related to well known effects of the response function during aging. We will discuss these results and their implication in greater details in section~\ref{sec:ccl}, after describing the results on the colloid gel. 

\section{ELECTRICAL NOISE OF LAPONITE}

We report in this section new results on electrical noise measurements in Laponite during the transition from a fluid like solution to a solid like colloidal glass. The main control parameter of this transition is the concentration of Laponite\cite{Laponite}, which is a synthetic clay consisting of discoid charged particles. It disperses rapidly in water to give gels even for very low mass fraction. Physical properties of this preparation evolve for a long time, even after the sol-gel transition, and have shown many similarities with standard glass aging\cite{Kroon,Bonn}. Recent experiments\cite{Bonn} have even proved that the structure function of Laponite at low concentration (less than $3 \%$ mass fraction) is close to that of a glass, suggesting the {\it colloidal glass} appellation.

In previous studies, we showed that the early stage of this transition was associated with a small aging of its bulk electrical conductivity, in contrast with a large variation in the noise spectrum at low frequency. As a consequence, the FDT in this material appeared to be strongly violated at low frequency in young samples, and it is only fulfilled for high frequencies and long times\cite{Bellon,BellonD,Buisson}. As in polycarbonate, this effect was shown to arise from a strong intermittency in the electrical noise of the samples, characterized by a strong deviation to a standard gaussian noise\cite{Buisson}. We summarize these results in the first part of this section, before presenting preliminary results on the role of concentration and of the DC polarization in the noise behavior.

\subsection{Experimental setup}

The experimental setup is similar to that of previous experiments\cite{Bellon,BellonD,Buisson}. The Laponite\cite{Laponite} dispersion is used as a conductive material between the two golden coated electrodes of a cell. It is prepared in a clean $\mathrm{N_2}$ atmosphere to avoid $\mathrm{CO_2}$ and $\mathrm{O_2}$ contamination, which perturbs the aging of the preparation and the electrical measurements. Laponite particles are dispersed at a concentration of $2.5 \%$ to $3 \%$ mass fraction in pure water under vigorous stirring for $300\,s$. To avoid the existence of any initial structure in the sol, we pass the solution through a $1\,\mu m$ filter when filling the cell. This instant defines the origin of the aging time $t_w$ (the filling of the cell takes roughly two minutes, which can be considered the maximum inaccuracy of $t_w$). The sample is then sealed so that no pollution or evaporation of the solvent can occur. At these concentrations, light scattering experiments show that Laponite\cite{Laponite} structure functions are still evolving several hundreds hours after the preparation, and that solid like structures are only visible after $100\,h$\cite{Kroon}. We only study the beginning of this glass formation process.

\begin{figure} \begin{center}

 \null  \hspace{10mm}  (a) \hspace{46mm} (b)  \hspace{53mm}  (c)  \null
 
\includegraphics[scale=0.9]{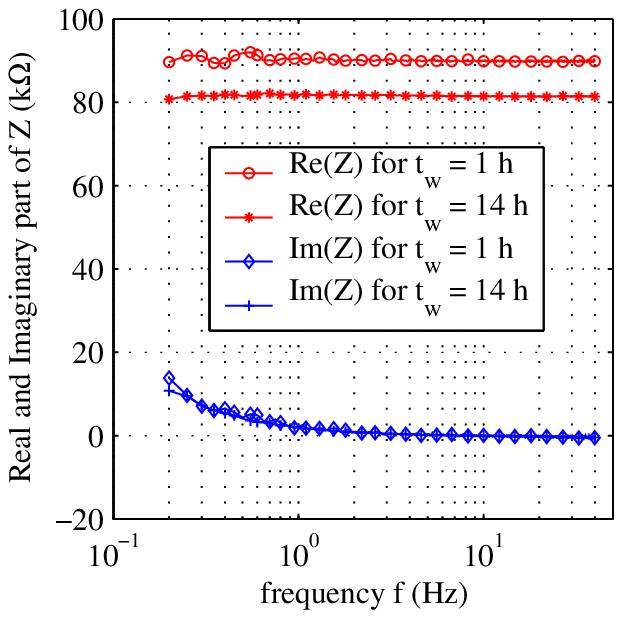}
\hfill
\includegraphics[scale=0.9]{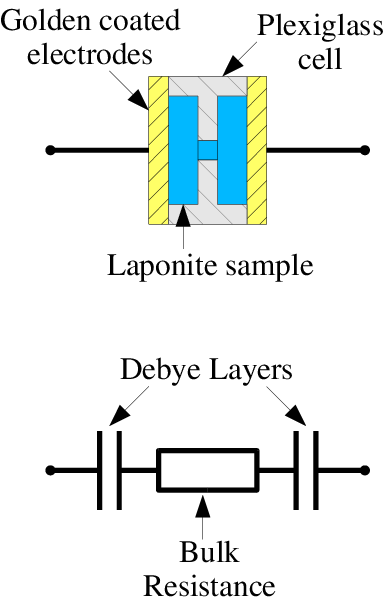}
\hfill
\includegraphics[scale=0.9]{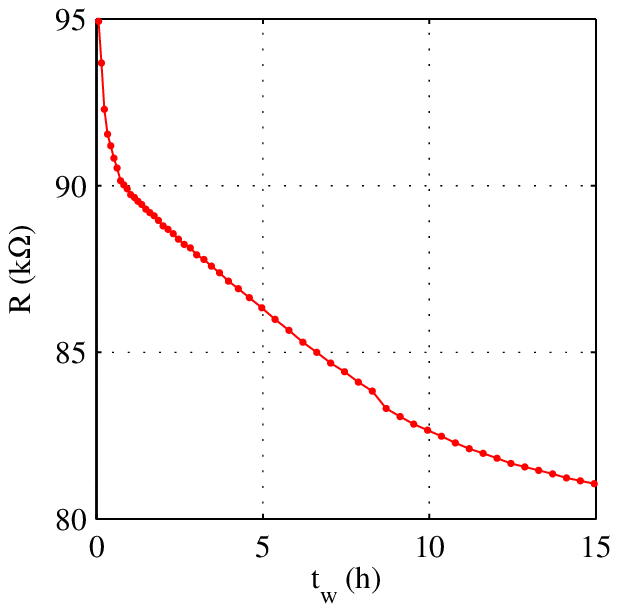}

 \end{center}
 \caption{{\bf Impedance of a $2.5\,wt\%$ Laponite cell. }(a) Frequency dependance of a sample impedance for 2 different aging times ($t_w=1\,h$ and $t_w=14\,h$). (b) Cell design and equivalent electrical model (c) Time evolution of the bulk resistance: this long time evolution is the signature of the aging of the colloidal suspension. In spite of the decreasing mobility of Laponite particles in solution during the formation of the gel, the electrical conductivity increases.}
 \label{fig:impedance}
\end{figure}

The two electrodes of the cell are connected to our measurement system, which records either the impedance value or the voltage noise across it. The electrical impedance of the sample is the sum of two effects: the bulk is purely conductive, the ions of the solution following the forcing field, whereas the interfaces between the solution and the electrodes give mainly a capacitive effect due to the presence of the Debye layers\cite{hunter}. This behavior has been validated using a four-electrode potentiostatic technique\cite{electrochem} to make sure that the capacitive effect is only due to the surface. In order to probe mainly bulk properties, the geometry of the cell is tuned to push the surface contribution to low frequencies: the cell consists in two large reservoirs where the fluid is in contact with the electrodes (area of $25\,cm^2$), connected through a small rigid tube --- see Fig.~\ref{fig:impedance}(b). The main contribution to the electrical resistance of the cell is given by the Laponite sol contained in this tube connecting the two tanks. Thus by changing the length and the section of this tube the total bulk resistance of the sample can be tuned around $R_{opt} = 100\,k\Omega$, which optimizes the signal to noise ratio of voltage fluctuations measurements with our amplifier. The cut-off frequency of the equivalent R-C circuit (composed by the series of the Debye layers plus the bulk resistance) is about $20 \,mHz$. In other words above this frequency the imaginary part of the cell impedance is about zero, as shown in Fig.~\ref{fig:impedance}(a). The time evolution of the resistance of one of our sample is plotted in Fig.~\ref{fig:impedance}(c): it is still decaying in a non trivial way after $24 h$, showing that the sample has not reached any equilibrium yet. This aging is consistent with that observed in light scattering experiments\cite{Kroon}.

\subsection{AC noise measurements in Laponite}

\begin{figure}[tb]
\begin{center}
 \includegraphics[width=8cm]{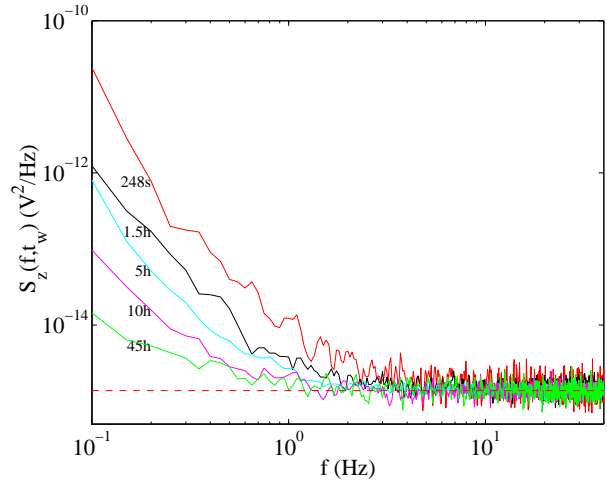}
 \end{center}
\caption{{\bf Voltage noise density for a $2.5\,wt\%$ Laponite sample. }The power spectrum density of voltage fluctuations across the impedance of a $2.5\,wt\%$ Laponite cell exhibit strong aging, and match the Nyquist formula prediction (horizontal dashed line) only for long times or large frequencies.}
\label{fig:LaponitePSD}
\end{figure}

In order to study the voltage fluctuations across the Laponite cell, we use a custom ultra low noise amplifier to raise the signal level before acquisition. To bypass any offset problems during this strong amplification process, passive high pass filtering above $30\,mHz$ is applied. The power spectrum density of the voltage noise of a $2.5\,wt\%$ Laponite preparation is shown in Fig.~\ref{fig:LaponitePSD}. As the dissipative part of the impedance $Re(Z)$ is weakly time and frequency dependent, one would expect from the Nyquist formula\cite{Nyquist} that so does the voltage noise density $S_{Z}$. But as shown in Fig.~\ref{fig:LaponitePSD}, we have a large deviation from this prediction for the lowest frequencies and earliest times of our experiment: $S_{Z}$ changes by several orders of magnitude between highest values and the high frequency tail. For long times and high frequencies, the FDT holds and the voltage noise density is that predicted from the Nyquist formula for a pure resistance at room temperature ($300\,K$). In order to be sure that the observed excess noise is not due to an artifact of the experimental procedure, we filled the cell with an electrolyte solution with a {\it p}H close to that of the Laponite preparation such that the electrical impedance of the cell was the same (specifically: $NaOH$ solution in water at a concentration of $10^{-3} \, mol \cdot l^{-1}$). In this case, the noise spectrum was flat and in perfect agreement with the Nyquist formula\cite{Bellon}.

\begin{figure}[tb]
 \begin{center}
 \includegraphics[width=7cm]{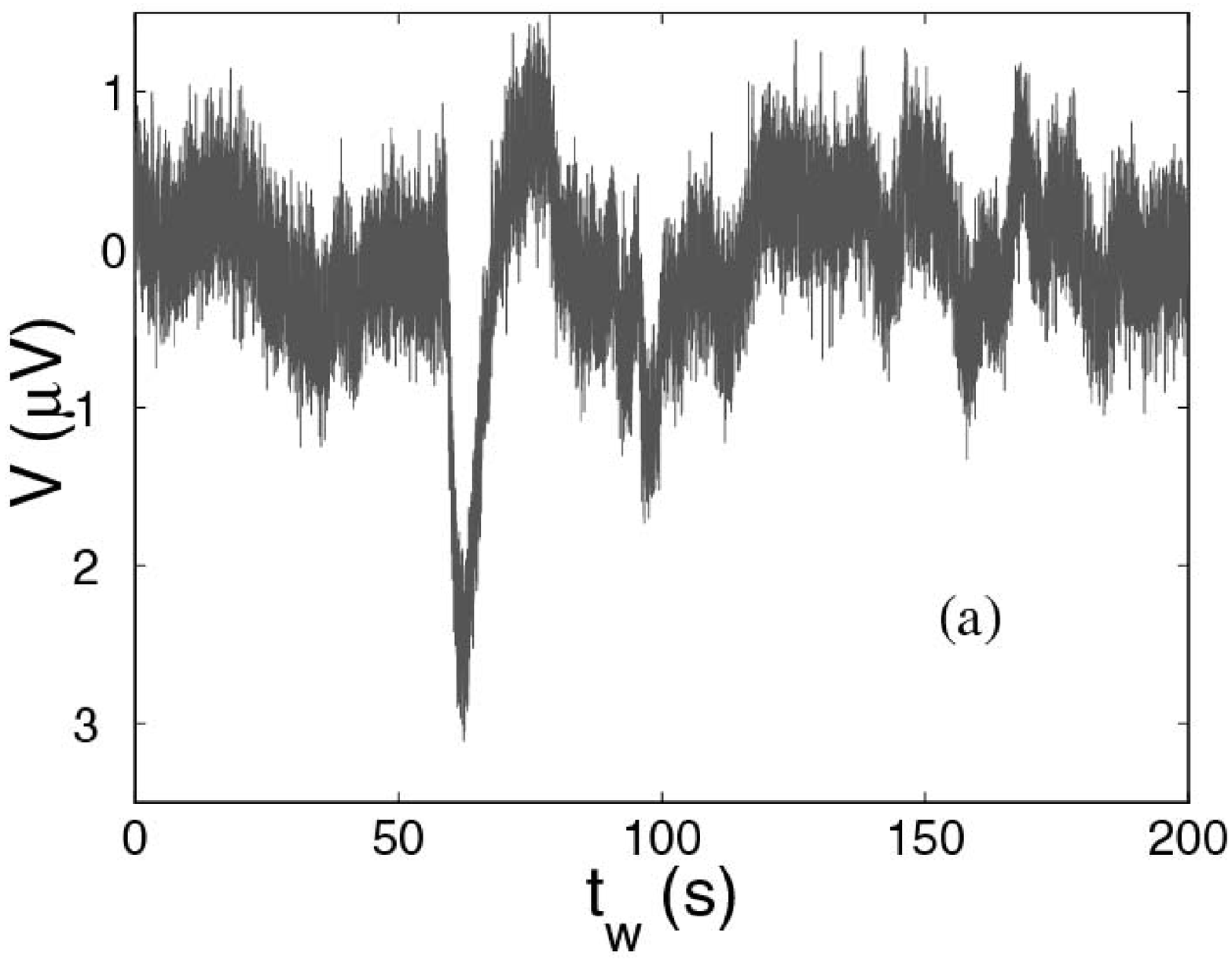}
 \hspace{15mm}
 \includegraphics[width=7cm]{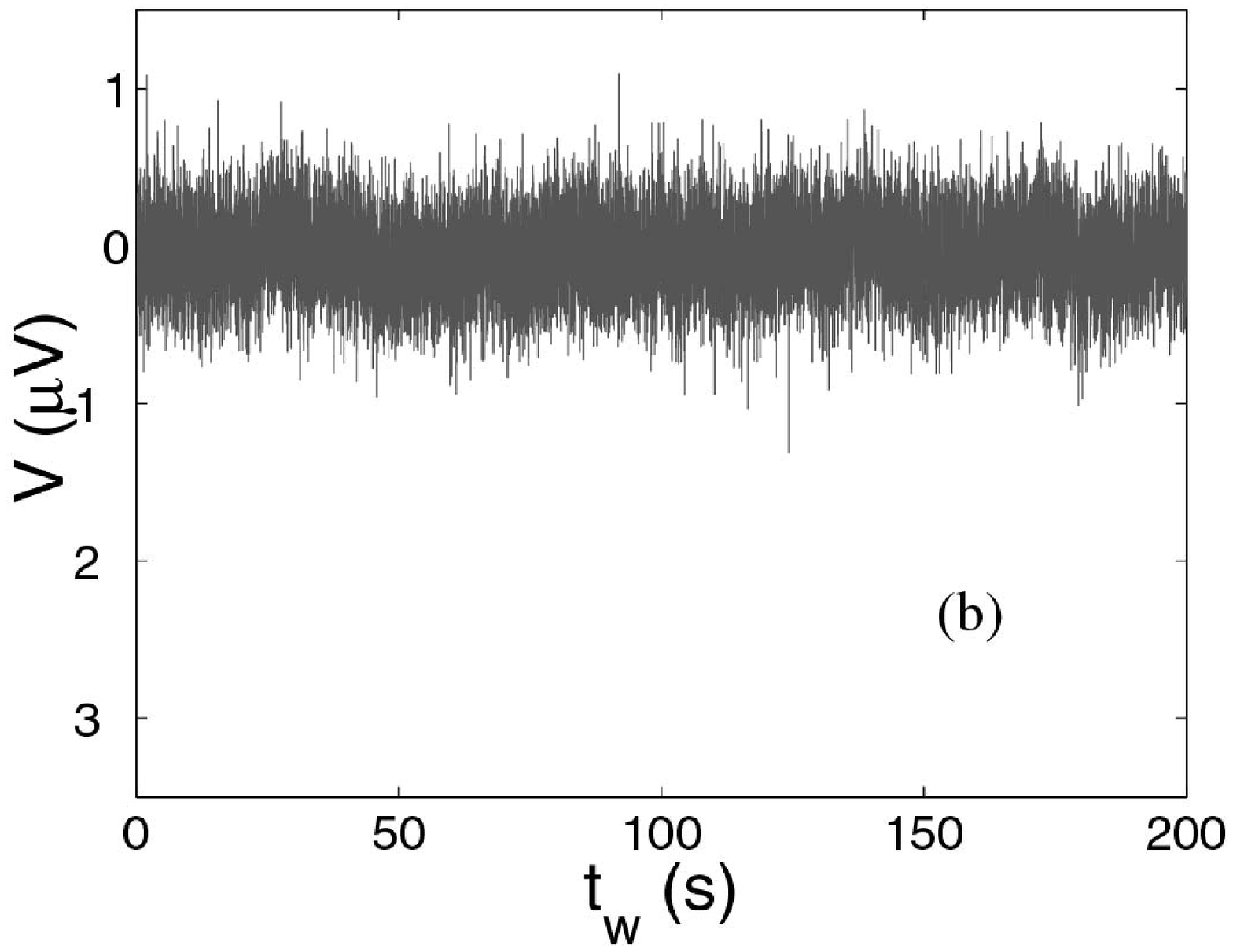}
 \end{center}
\caption{{\bf Voltage noise signal in a $2.5\,wt\%$ Laponite sample. }(a) Noise
signal, 2 hours after the Laponite preparation, when FDT is
violated. (b) Typical noise signal when FDT is not violated ($t_w=50\,h$). }
\label{fig:signal2.5}
\end{figure}

Aiming at a better understanding of the physics underlaying such a behavior, we have directly analyzed the voltage noise across the Laponite cell. This test can be safely done in our experimental configuration as the amplifier noise is negligible with respect to the voltage fluctuations across cell, even for the lowest levels of the signal, that is when the FDT is satisfied. In Fig.~\ref{fig:signal2.5}(a) we plot a typical signal measured $2h$ after the gel preparation, when the FDT is strongly violated. The signal plotted in Fig.~\ref{fig:signal2.5}(b) has been measured when the system has relaxed and FDT is satisfied in all the frequency range. By comparing the two signals we immediately realize that there are important differences. The signal in Fig.~\ref{fig:signal2.5}(a) is interrupted by bursts of large amplitude which are responsible for the increasing of the noise in the low frequency spectra (see Fig.~\ref{fig:LaponitePSD}). The relaxation time of the bursts has no particular meaning, because it corresponds just to the characteristic time of the filter used to eliminate the very low frequency trends. As time goes on, the amplitude of the bursts reduces and the time between two consecutive bursts becomes longer and longer. Finally they
disappear as can be seen in the signal of Fig.~\ref{fig:signal2.5}(b) recorded for a $50\,h$ old preparation, when the system satisfies FDT.

\begin{figure}[tb]
 \begin{center}
 \includegraphics[width=8cm]{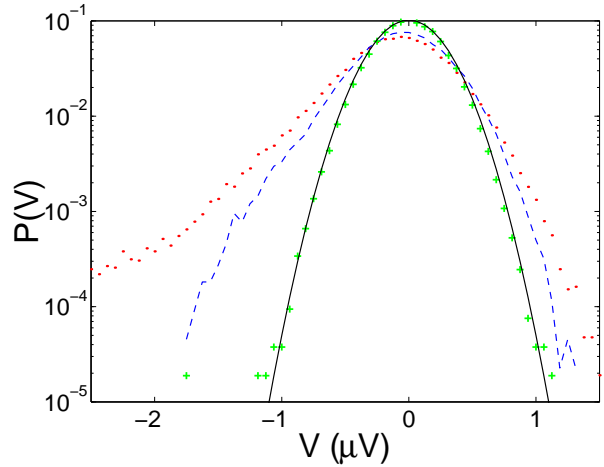}
 \end{center}
\caption{{\bf PDF of the voltage noise in a $2.5\,wt\%$ Laponite sample. }Typical PDF
of the noise signal at different times after preparation, with
from top to bottom: $
 \ (...) t_w=1\,h, \ (- -) t_w=2\,h, \ (+)
t_w=50\,h$. The continuous line is obtained from the FDT prediction.
} \label{fig:PDFlaponite2.5}
\end{figure}

As in polycarbonate, the intermittent properties of the noise can be characterized by the PDF of the voltage fluctuations. To compute these distributions, the time series are divided in several time windows and the PDF are computed in each of these window. Afterwards the result of several experiments are averaged. The distributions computed at different times are plotted in Fig.~\ref{fig:PDFlaponite2.5}. We see that at short $t_w$ the PDF presents heavy tails which slowly disappear at longer $t_w$. Finally a Gaussian shape is recovered after $t_w=16\,h $. This kind of evolution of the PDF clearly indicate that the signal is very intermittent for a young sample and it relaxes to the Gaussian noise at long times.

\subsection{Influence of concentration}

\begin{figure}
\begin{center}
\begin{minipage}[b]{0.48\linewidth}
\begin{center}

\includegraphics[width=7.02cm]{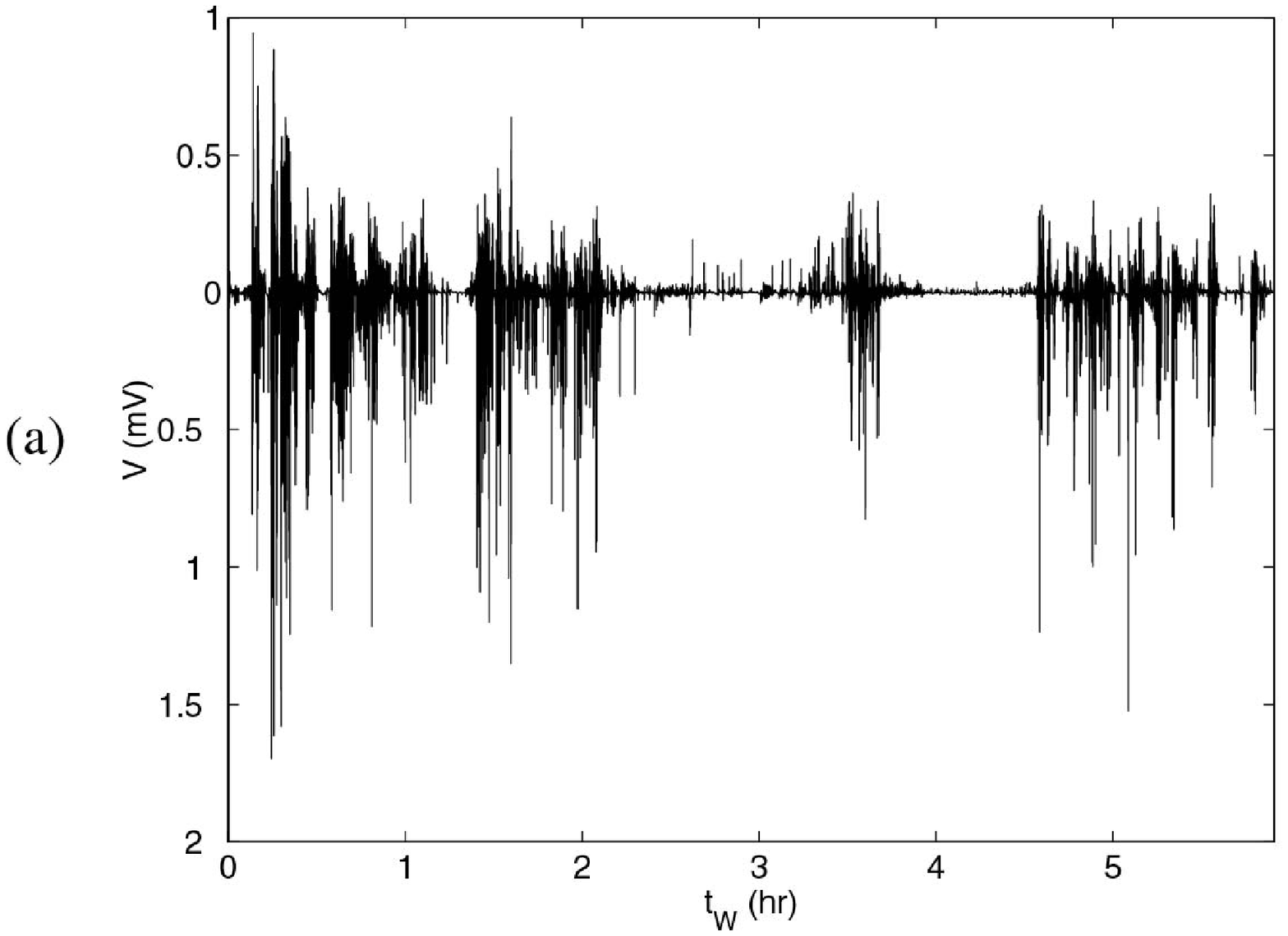}

\vspace{0.5cm}

\includegraphics[width=7.02cm]{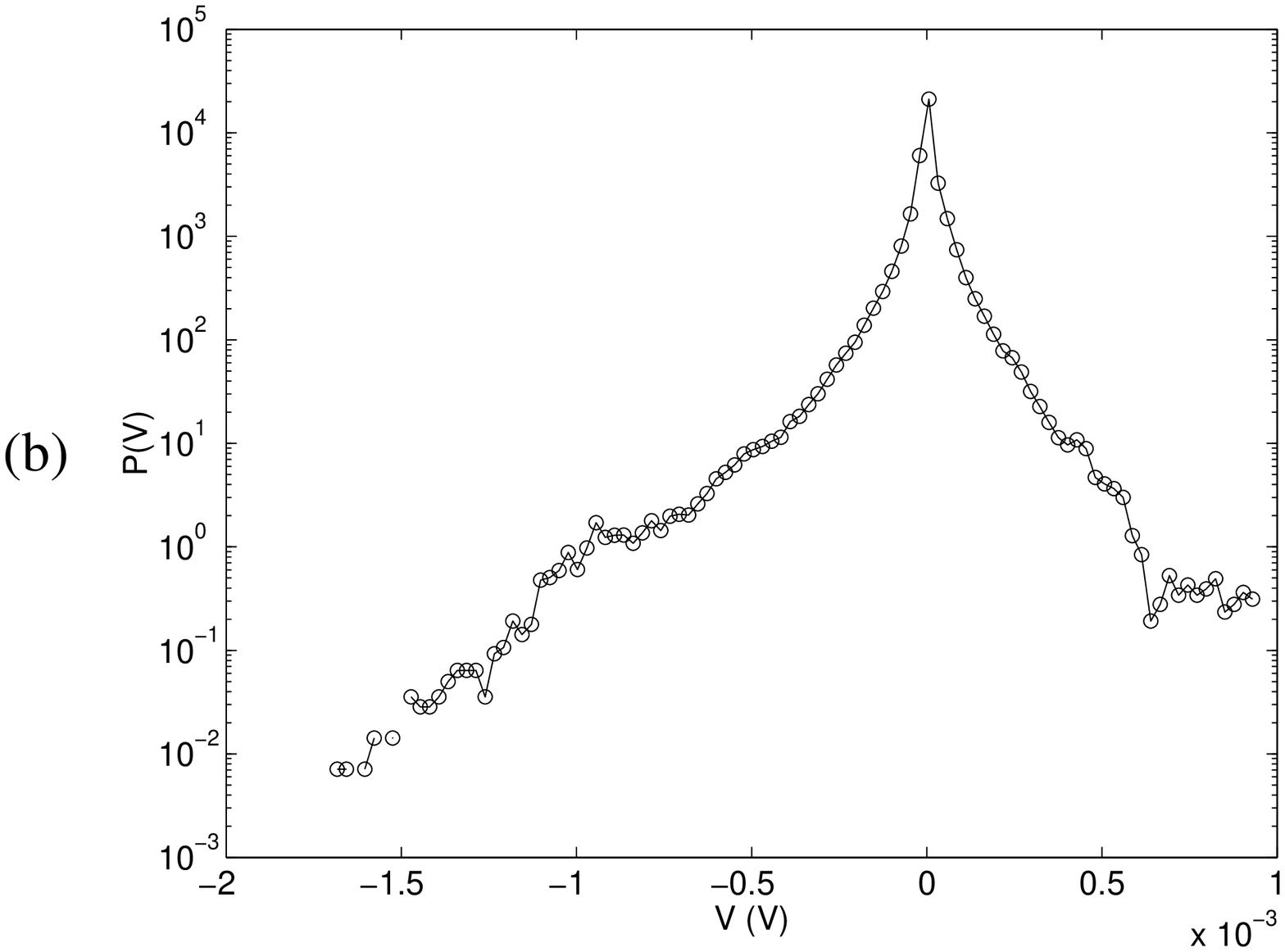}

\end{center}
\caption{{\bf Voltage noise in a $3\,wt\%$ Laponite sample. }(a) During the first hours, the voltage noise is dominated by huge intermittent fluctuations: bursts over $1\,mV $ are detected, when thermal noise should present a typical $1\,\mu V$ rms amplitude. (b) The PDF of this intermittent signal departs clearly from a gaussian distribution. \label{fig:Laponite3Filtered}}

\vspace{1cm}

\begin{center}
\includegraphics[width=5.85cm]{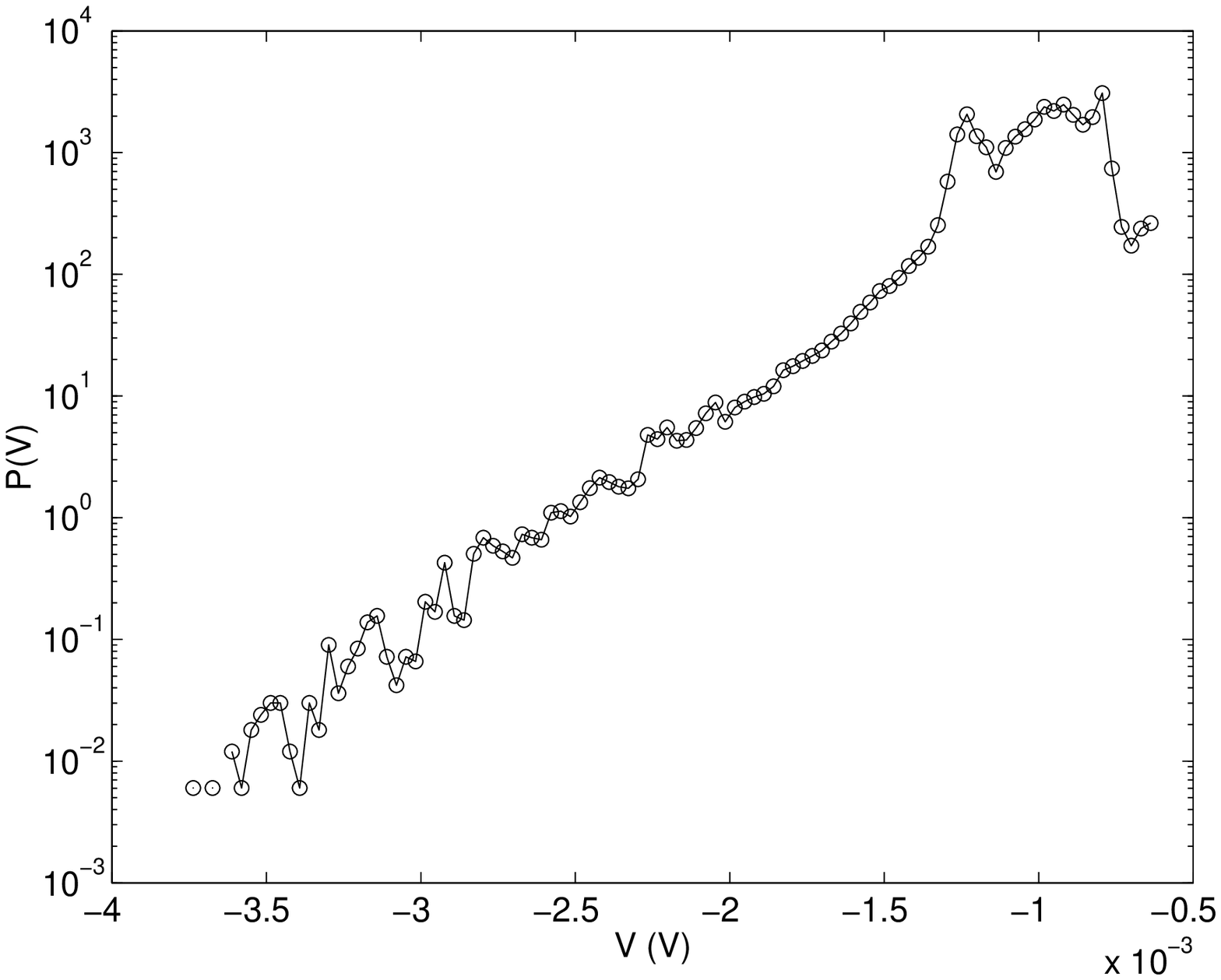}
\end{center}
\caption{{\bf PDF of the unfiltered voltage noise in Laponite. }PDF of the signal plotted in Fig.~\ref{fig:DirectTot}(a), corresponding to the first $6\,h$ of a $3\,wt\%$ Laponite sample. \label{fig:DirectPDF}}
\end{minipage}
\hfill
\begin{minipage}[b]{0.48\linewidth}
\begin{center}
\includegraphics[width=7.2cm]{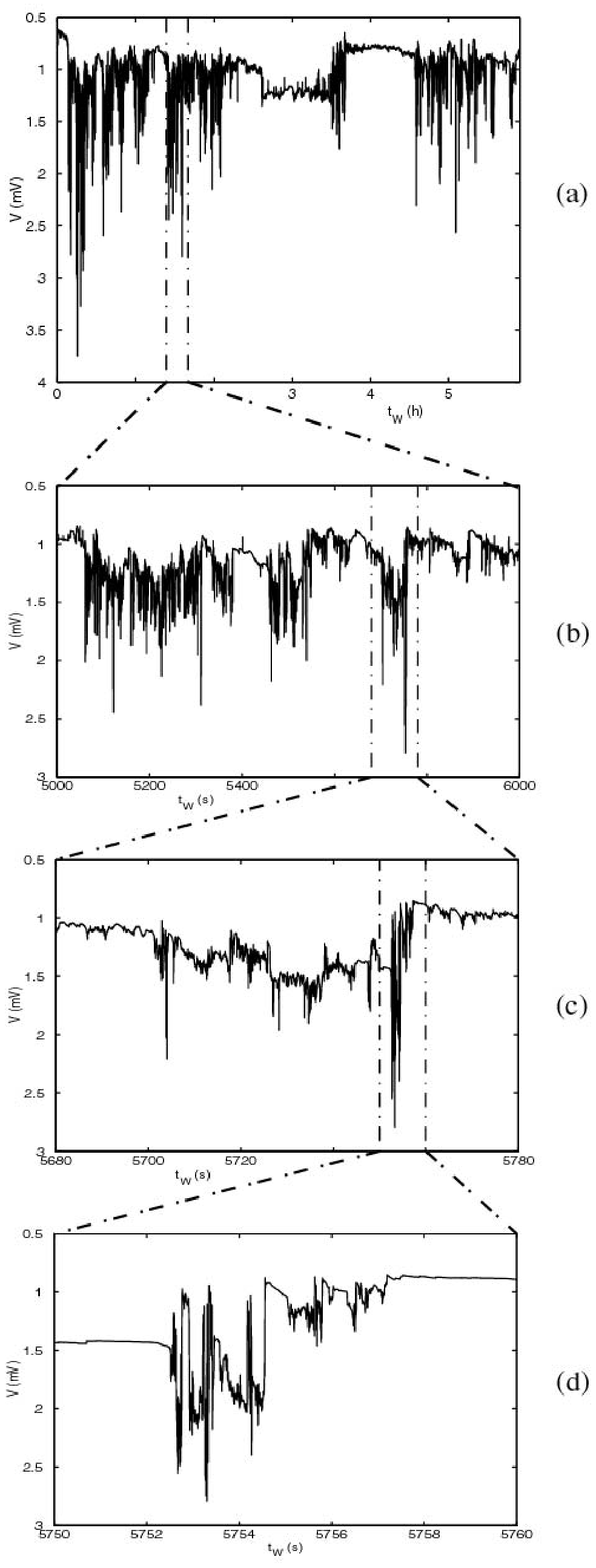}
\end{center}
\caption{{\bf Unfiltered voltage noise in Laponite. }(a) Voltage fluctuations recorded during the first $6\,h$ of a $3\,wt\%$ Laponite sample, corresponding to the filtered signal of Fig.~\ref{fig:Laponite3Filtered}(a). The intermittency is much more pronounced on the direct signal, and is associated to discrete steps in polarization. (b) to (d) successive blowups of the signal, showing a rich behavior with nested structures down to $0.1\,s$, the shortest time scale resolved here. \label{fig:DirectTot}}
\end{minipage}
\end{center}
\end{figure}

To check for the influence of concentration on these results, we recently started new series of measurements with $3\,wt\%$ Laponite preparations. In Fig.~\ref{fig:Laponite3Filtered}(a) we plot a typical signal measured during the first 6 hours of such a sample. Again, this signal is interrupted by bursts of very large amplitude. As time goes on, the amplitude of the bursts reduces and the time between two consecutive bursts becomes longer and longer. Finally they disappear after a few days, and we only observe classic thermal noise. The main difference with less concentrated samples is in the amplitude and density of this intermittency: now bursts over $1\,mV $ are detected, when thermal noise should present a typical $1\,\mu V$ rms amplitude, and they are much more frequent. This difference is also clear on the PDF of the signal, plotted in Fig.~\ref{fig:Laponite3Filtered}(b). The non gaussian shape is much more pronounced, and the presence of heavy tails clearly indicate that the signal is very intermittent at the beginning of the experiment. In fact, the dynamic is so important that we don't even have enough precision to resolve the classic thermal fluctuations predicted by the Nyquist formula in this measurement. The influence of increasing the concentration of Laponite preparation thus appears to be somehow similar to the effect of increasing the cooling rate during the quench of the polymer glass: the resulting dynamics is more intermittent in both cases.

\subsection{Influence of DC polarization}

One of the most puzzling features of the voltage noise measurements, beyond their large amplitudes, is the marked asymmetry of the intermittent fluctuations, which is well apparent in Fig.~\ref{fig:PDFlaponite2.5} and \ref{fig:Laponite3Filtered}. There is indeed no a priori justification for the negative fluctuations to be predominant. In order to gain a deeper insight into the nature of these fluctuations, we recently managed to perform some measurements removing the filters that were used to eliminate the very low frequency trends but could mask some details of the intermittent dynamics as well as the presence of steps in the fluctuating signal.

One major difficulty in this measurement lies in the presence of a spontaneous polarization in the electrolytic cell, which is due to an asymmetry between the Debye layers on the electrodes. In fact, an initial voltage up to some tens of millivolts is observed between the electrodes after filling the cell with the Laponite preparation, then it slowly decreases eventually reaching an ``equilibrium'' value of about $1\,mV$ after some hours. The filters were initially introduced in order to reduce the dynamics of the signal and to be able of resolving the intermittent dynamics as well as the Gaussian noise. In fact, we found out that the polarization can be rapidly decreased near the $1\,mV$ value by short-circuiting the electrodes for some minutes prior to the noise measurement. In this condition the signal is centered about $1\,mV$ and the intermittencies are also of the order of $1\,mV$, while the Gaussian noise has about $1\,\mu V$ rms amplitude. 

To compare the difference between the direct and filtered signal, we plot in Fig.~\ref{fig:DirectTot}(a) the direct signal corresponding to the filtered one of Fig.~\ref{fig:Laponite3Filtered}(a), that is during a single experiment on the first 6 hours of a $3\,wt\%$ Laponite sample. More precisely, the measurement in Fig.~\ref{fig:DirectTot}(a) is obtained with a x400 preamplifier, while the measurement in Fig.~\ref{fig:Laponite3Filtered}(a) is subject to a second stage x10 amplification preceded by a high pass filter with cutoff frequency $30\,mHz$. The results in the figures are expressed in term of input voltage units.

Three main differences in relation to the filtered configuration are immediately apparent: (1) the presence of the polarization bias at about $-1\,mV$; (2) the asymmetry in the intermittent fluctuations is much more pronounced; (3) the noise turns out not to be stationary, but rather presents some steps associated with the intermittencies. Moreover, if we take a closer look at the unfiltered signal, the intermittencies reveal a much richer behavior with nested structures at all time scales down to the shortest resolved time scale of $0.1\,s$ (this limit is due to the low-pass filtering of the signal prior to discretization). This behavior is illustrated in a series of blowups of the unfiltered signal in Fig.~\ref{fig:DirectTot}(a) to (d).

If we now try to analyze this intermittent behavior using the PDF of the unfiltered voltage fluctuations, we observe a rather different behavior from that of the filtered signal (see Fig.~\ref{fig:DirectPDF} and \ref{fig:Laponite3Filtered}(b)). The following features can be noticed: (1) the asymmetry is much more pronounced; (2) the left tail has a clear exponential decay; (3) the peak region is broadened by the fluctuations of the mean of the signal. This region clearly needs further extensive ensemble averaging to be well defined and analyzed precisely.

These preliminary measurements involving the direct voltage noise signal show that polarization effects are important and should be considered to analyze in depth the intriguing features of electrical noise in Laponite preparation. Other experiments will anyway be necessary to probe the Debye layers and their behavior during the aging process, in order to have a better understanding of the whole system.

\section{DISCUSSION AND CONCLUSIONS}
\label{sec:ccl}
 
In the previous sections of this paper we have presented the measurements of the electrical thermal noise during the aging of two very different materials: a polymer and a colloidal suspension. The main results of these experiments are:
\begin{itemize}
\item[$\bullet$]  At the very beginning of aging   the noise amplitude for both materials  is much  larger than what  predicted by Nyquist relations. In other words Nyquist relations, or generally FDT, are violated because the material are out of equilibrium: they are aging.
\item[$\bullet$] The noise slowly relaxes to the usual value after a very long time.
\item[$\bullet$] For the polymer there is a large difference between fast and slow quenches. In the first case the thermal signal is strongly intermittent, in the second case this feature almost disappears.
\item[$\bullet$] The colloidal suspension signal is strongly intermittent, all the more as concentration is increased. The asymmetry of the noise may be linked to the spontaneous polarization of the cell.
 \end{itemize}

We want first to discuss the intermittence of the signal, which has been observed in other aging systems. Our observations are reminiscent of the intermittence observed in the local measurements of polymer dielectric properties\cite{Israeloff_Nature} and in the slow relaxation dynamics of another colloidal gel\cite{Mazoyer}. Indeed several theoretical models predict an intermittent dynamics for an aging system. For example the trap model\cite{trap} which is based on a phase space description of the aging dynamics. Its basic ingredient is an activation process and aging is associated to the fact that deeper and deeper valleys are reached as the system evolves\cite{bertin,Sollich,Miguel}. The dynamics in this model has to be intermittent because either nothing moves or there is a jump between two traps\cite{Sollich}. This contrasts, for example, with mean field dynamics which is continuous in time\cite{Kurchan}. Furthermore two very recent theoretical models predict skewed PDF both for local\cite{Crisanti} and global variables\cite{SibaniI}. This is a very important observation, because it is worth to notice that one could expect to find intermittency in local variable but not in the global. Indeed in global measurements, fluctuations could be smoothed by the volume average and therefore the PDF  should be  Gaussian. This is not the case both for our data and for the numerical simulations of aging models\cite{SibaniI}. In order to push the comparisons with these models of intermittency on a more quantitative level one should analyze more carefully the PDF of the time between events, which is very different in the various models\cite{trap,SibaniII}. Our statistics is not yet enough accurate to give clear answers on this point, thus more measurements are necessary to improve the comparisons between theory and experiment.

Going back to the analysis of our experimental data there is another observation, which merits to be discussed. This concerns the difference between fast and slow quenches in polycarbonate.  This result  can be understood considering that  during the fast quench the material is frozen in a state which is a highly out of equilibrium one for the new temperature. This is not the case for a slow quench. More precisely one may assume that when an aging system is quenched very fast, it explores regions of its phase space that are completely different than those explored in the quasi-equilibrium states  of a slow quench. This  assumption is actually supported by two recent theoretical results\cite{Mossa,BertinII}, which were obtained in order to give a satisfactory explanation, in the framework of the more recent models, of the old Kovacs effect on the volume expansion\cite{Kovacs}. Thus our results based on the noise measurements can be interpreted in the same way.

Finally, we want to discuss the analogy between the electrical thermal noise  in the fast quench experiment of the polymer and that of the gel during the sol-gel transition. In spite of the physical mechanisms that are certainly very different, the statistical properties of the signals are very similar. Thus one may ask, what is the relationship between the fast quench in the polymer and the sol-gel transition. As we already mentioned, during the fast quench the polymer is strongly out of equilibrium which is the same situation for the liquid-like state at the very beginning of the sol-gel transition. The speed of the sol-gel transition is controlled in Laponite by the initial Laponite/water concentration and therefore intermittency should be a function of this parameter. Preliminary measurements seem to confirm this guess: the higher the concentration, the stronger the intermittency.

In conclusion we have shown that the experimental study of thermal noise associated with that of the response function gives new insights on the interpretation of the aging dynamics. The important question is now to understand the physical origin of these big events in the electrical thermal noise of the sample. Mechanical\cite{BellonD}, acoustical and light scattering measurements performed in parallel with dielectric measurements could clarify this problem. Indeed the analysis of electrical noise alone is certainly not enough and measurements on mechanical noise will be certainly important in order to verify whether this kind of behavior pertains only to electrical variables or it is more general.  More measurements are required to give more quantitative answers but this new analysis of aging systems  is certainly useful to improve the quantitative comparisons between aging models and experiments.

\acknowledgments

We acknowledge useful discussion with J. Kurchan, J.P.
Bouchaud, F. Ritort and P. Sibani. We thank P. Metz for technical support.
This work has been partially supported by the R\'egion
Rh\^one-Alpes contract ``Programme Th\'ematique: Vieillissement
des mat\'eriaux amorphes''.

\end{document}